\documentclass[a4paper,11pt]{article}
\pdfoutput=1 

\usepackage{jcappub} 

\makeatletter

\patchcmd{\maketitle}
  {\Large \bfseries\raggedright\sffamily\the\auth@toks}
  {\large \bfseries\raggedright\sffamily\the\auth@toks}
  {}{\PackageWarning{paper5}{Author font patch failed}}
\patchcmd{\maketitle}
  {\the\affil@toks}
  {\small\the\affil@toks}
  {}{\PackageWarning{paper5}{Affiliation font patch failed}}
\patchcmd{\maketitle}
  {E-mail: \the\email@toks}
  {\footnotesize E-mail: \the\email@toks}
  {}{\PackageWarning{paper5}{Email font patch failed}}

\usepackage{amsmath}
\usepackage{booktabs}
\usepackage{threeparttable}
\usepackage{tabularx}
\usepackage{mathrsfs}

\usepackage{listings}

\usepackage[T1]{fontenc} 

\title{LIMFAST. V. Probing Reionization with the Synergy Between 21\,cm and Photometric Galaxy Surveys}

\author[a,*]{Guochao Sun,\note[*]{Corresponding author.}}
\author[b]{Adam Lidz,}
\author[c,d]{Tzu-Ching Chang,}
\author[a]{Claude-Andr\'{e} Faucher-Gigu\`{e}re,}
\author[e]{Steven R. Furlanetto,}
\author[b]{Aritra Kundu,}
\author[c]{and Jordan Mirocha}

\apptocmd{\afterAuthorSpace}{\vspace{0.2cm}}{}{}

\affiliation[a]{\textit{CIERA and Department of Physics and Astronomy, Northwestern University, 1800 Sherman Ave, Evanston, IL 60201, USA}}
\affiliation[b]{\textit{University of Pennsylvania, Department of Physics and Astronomy, 209 S. 33rd Street, Philadelphia, PA 19104, USA}}
\affiliation[c]{\textit{Jet Propulsion Laboratory, California Institute of Technology, 4800 Oak Grove Drive, Pasadena, CA 91109, USA}}
\affiliation[d]{\textit{California Institute of Technology, 1200 E. California Blvd., Pasadena, CA 91125, USA}}
\affiliation[e]{\textit{Department of Physics and Astronomy, University of California, Los Angeles, CA 90095, USA}}

\apptocmd{\afterAffiliationSpace}{\vspace{0.2cm}}{}{}

\emailAdd{guochao.sun@northwestern.edu}
\emailAdd{alidz@sas.upenn.edu}
\emailAdd{tzu-ching.chang@jpl.nasa.gov}
\emailAdd{cgiguere@northwestern.edu}
\emailAdd{sfurlane@astro.ucla.edu}
\emailAdd{aritra@sas.upenn.edu}
\emailAdd{mirocha@caltech.edu}

\apptocmd{\afterEmailSpace}{\vspace{0.3cm}}{}{}

\abstract{Cross-correlating the 21\,cm signal from intergalactic neutral hydrogen with the galaxy distribution is a powerful probe of the Epoch of Reionization and the physical mechanisms driving it. While upcoming surveys with the SKA-Low radio telescope and the Roman Space Telescope offer ideal synergies for this measurement, foreground mitigation of 21\,cm data requires exquisite redshift precision of galaxies, effectively excluding the rich photometrically selected samples of Lyman-break galaxies (LBGs) with typical redshift uncertainties $\sigma_z \gtrsim 0.1$. We overcome this challenge using a squared statistic of the foreground-cleaned 21\,cm signal. Squaring the filtered 21\,cm signal couples surviving high $k_\parallel$ modes into low $k_\parallel$ fluctuations accessible to photometric LBG surveys. This higher-order cross-correlation statistic remains measurable even when direct cross-correlations are strongly suppressed. Applying Fisher matrix analysis to this LBG--21\,cm$^2$ cross-correlation signal simulated by the \texttt{LIMFAST} code over $6 \lesssim z \lesssim 10$, we investigate the prospects for constraining EoR physics combining SKA-Low 21\,cm measurements and Roman photometric galaxy samples. For an overlapping survey area covering 10\,deg$^2$, we find that the LBG--21\,cm$^2$ cross-correlation is detectable at high significance across much of the EoR through 1000\,hours of SKA-Low integration with 15\,MHz bandwidth and a Roman photometric galaxy sample from the High-Latitude Wide-Area Survey (HLWAS) reaching $m_\mathrm{AB,lim}=27$ and photo-$z$ uncertainties of $\sigma_z \sim 0.1$--0.3. The characteristic sign evolution of the large-scale cross-power spectrum---driven primarily, though not exclusively, by the evolving neutral fraction---provides strong constraints on the physical parameters governing the ionization and thermal history of the IGM. Our approach unlocks the promising potential of the abundant LBG population for EoR cross-correlation studies with the 21\,cm signal.}


\begin{document}

\maketitle
\flushbottom

\section{Introduction}
\label{sec:intro}

The Epoch of Reionization (EoR) marks a major phase transition of the intergalactic medium (IGM) from a predominantly cold and neutral state into the hot and highly ionized medium observed at later times. It encodes a wealth of information about the first generations of galaxies and the physical processes through which they heated and ionized their surroundings \citep{BarkanaLoeb2001,GnedinMadau2022}. Among the most promising probes of the EoR is the redshifted 21\,cm line of neutral hydrogen, which directly traces the evolving density, ionization, and thermal state of the IGM \citep{Furlanetto2006,PritchardLoeb2012,Shimabukuro2023}. On the other hand, high-redshift galaxy surveys provide a complementary view by identifying and characterizing the sources of ionizing and heating radiation \citep{Fan2006,Robertson2022}. Cross-correlating these observables can therefore constrain the timing and topology of reionization and connect the evolving IGM to the source populations \citep{FurlanettoLidz2007,Lidz2009,Sobacchi2016,Kubota2018,LaPlante2023,Moriwaki2024,Gagnon-Hartman2025,HutterHeneka2026}. In particular, the 21\,cm--galaxy cross-correlation is found to be sensitive to the changing balance among density, spin temperature, and ionization fluctuations and can change sign as the dominant physical process evolves from heating to ionization \citep{Moriwaki2024}.

A major practical challenge, however, is that realistic 21\,cm foreground mitigation removes severely contaminated long-wavelength modes along the line-of-sight (LOS) direction, which precisely overlap with the scales probed by galaxy surveys with modest-quality photometric redshifts. Forecasts show that the accessible EoR window is jointly defined by the 21\,cm foreground wedge, galaxy redshift uncertainty, and survey area \citep{FurlanettoLidz2007,HutterHeneka2026}. In particular, assuming a moderate foreground wedge, photometric redshift uncertainties ($\sigma_z \gtrsim 0.1$) can effectively eliminate the shared LOS modes required for a useful cross-correlation detection, leaving the spectroscopic galaxy samples as the only viable path for measuring the direct cross-correlation \citep{Lidz2009,HutterHeneka2026}. This limitation is clearly undesirable: upcoming photometric imaging surveys of Lyman-break galaxies (LBGs) by, e.g., the recently launched Roman Space Telescope, will cover wide areas and redshift ranges at various depths, thus offering a holistic map of the EoR landscape. Requiring spectroscopic redshifts can significantly reduce the usable galaxy sample and restrict its redshift coverage through emission line sensitivities and instrumental bandwidth. For example, Roman grism surveys of EoR galaxies depend heavily on Ly$\alpha$ emission \citep{Wold2024}, which falls into the 1--1.92\,$\mu$m bandpass of the grism only at $z \gtrsim 7.2$. Direct cross-correlations based on these Roman Ly$\alpha$-selected samples would thus be limited to the early stages of reionization. Moreover, the accuracy of redshifts derived from slitless spectroscopy depends critically on controlling contamination and confusion from the dispersed traces of neighboring objects \citep{Brammer2012,Xiao2026}, especially when observing faint high-$z$ galaxies in crowded fields or when only one single emission line is detected. Developing alternative statistics that combine the abundant photometrically selected galaxies with foreground-filtered 21\,cm measurements would therefore substantially expand the available cross-correlation science for a deeper understanding of the EoR.

In this paper, we investigate such a statistic by cross-correlating photometric galaxies with a filtered-then-squared 21\,cm field. While the direct cross-correlation probes only common Fourier modes of the two fields, squaring the 21\,cm field couples pairs of surviving high-$k_\parallel$ modes to another mode at low $k_\parallel$, where the photometric galaxy field retains most of its information. The resulting galaxy--21\,cm$^2$ cross-power spectrum is therefore a higher-order statistic formally equivalent to an integrated or projected cross-bispectrum, probing how the variance of small-scale 21\,cm modes are correlated with the large-scale galaxy field---information strongly suppressed for the direct two-point cross-correlation after foreground and photo-$z$ filtering. This construction is closely related to previous studies of post-reionization 21\,cm cross-bispectra involving CMB lensing or photometric galaxy fields \citep{Guandalin2022,Moodley2023,Noble2026}, including approaches designed to recover information suppressed by foreground cleaning. Our approach instead extends this idea to the EoR and compresses the three-point information into the cross-power spectrum of galaxies with a filtered-then-squared 21\,cm field, whose redshift evolution and scale dependence encode the evolving contributions of density, ionization, and spin temperature. In a related EoR application, we study the cross-correlation of filtered-then-squared 21\,cm maps with the cosmic near-infrared background \citep{Sun2025}. The present analysis does more than replace the NIRB with another projected field in that it models a discrete population of Roman-selected LBGs as ionizing sources, incorporates the Fourier space mismatch, and forecasts the detectability and constraining power for astrophysical parameters. Other recent applications include cross-correlations involving filtered-then-squared 21\,cm maps and the squared kSZ signal \citep{Zhou2025,Yuwen2026}. Early explorations of the position-dependent power spectrum between CMB lensing and the Ly$\alpha$ forest also employ a similar concept \citep{Doux2016}. 

We quantify this higher-order statistic using physically connected 21\,cm and photometric galaxy fields generated with the semi-numerical code \texttt{LIMFAST} \citep{MasRibas2023,Sun2023} anchored to halo catalogs from the mini-Uchuu simulations \citep{Ishiyama2021}. We construct Roman-like LBG samples reaching $m_{\mathrm{AB,lim}}=27$ and forecast their cross-correlation with SKA-Low observations during the EoR. For 1000\,hours of SKA-Low observations overlapping a $10\,\mathrm{deg}^{2}$ Roman deep survey, we find cumulative signal-to-noise ratios ranging from approximately 5 to 20 during the bulk of reionization depending on the redshift and photo-$z$ error. The large-scale cross-correlation changes sign as the relative importance of ionization, density, and thermal fluctuations evolves, providing complementary information at different redshifts. We then perform a Fisher analysis to demonstrate the resulting sensitivity to the ionizing escape fraction, X-ray heating efficiency, feedback-regulated star formation, and minimum halo mass for star formation. We also examine the physical origin of the sign evolution using a Morlet-wavelet analysis and a decomposition of 21\,cm fluctuations. Our analysis shows that Roman LBG samples can contribute meaningfully to EoR science through cross-correlations with the 21\,cm signal even without spectroscopic redshifts, thereby expanding the galaxy sample available for joint analyses. More broadly, this approach provides a general solution to accessing useful higher-order cross-correlation information between LIM signals contaminated by spectrally smooth foregrounds and tracers with limited radial resolution, including photo-$z$ galaxies and projected fields such as the CMB secondary anisotropies and CIB. It therefore complements other approaches for retaining or recovering cross-correlation information in the presence of foreground filtering, including tidal reconstruction of lost radial modes \citep{Zhu2018}, direct correlations enabled by mode coupling associated with evolution along the lightcone \citep{Shen2026}, and non-Gaussian $k$-nearest-neighbor statistics \citep{Chakraborty2026}.

The remainder of this paper is organized as follows. In section~\ref{sec:models}, we summarize the key physical and observational modeling ingredients for the galaxy and 21\,cm signals relevant to the cross-correlations studied here. In section~\ref{sec:sims}, we describe the \texttt{LIMFAST} framework used to simulate the target cross-correlation signals during the EoR, including new implementations designed to facilitate cross-correlation studies with galaxy surveys. We present our main results in section~\ref{sec:results}, and then discuss the limitations of the present analysis, possible extensions, and our conclusions in section~\ref{sec:conclusions}. Throughout this paper, we adopt the Planck cosmological parameters \citep{Planck2016}, consistent with those assumed in the Uchuu simulations \citep{Ishiyama2021}.

\section{Models} \label{sec:models}

\subsection{21\,cm signal of neutral hydrogen}

The redshifted cosmological 21\,cm signal in terms of the differential brightness temperature against the cosmic microwave background (CMB) can be modeled as,
\begin{equation}
\delta T_\mathrm{b} \propto x_\mathrm{HI} \left( 1 + \delta \right)
\left[1-\frac{T_{\rm CMB}}{T_S}\right]
\left(\frac{\Omega_\mathrm{b} h^2}{0.022}\right)
\left(\frac{0.14}{\Omega_\mathrm{m} h^2}\frac{1+z}{10}\right)^{1/2} \left[ 1 + \frac{1+z}{H(z)} \frac{\partial v_{\parallel}}{\partial r_{\parallel}} \right]^{-1},
\end{equation}
where $x_\mathrm{HI}$ is the neutral hydrogen fraction, $\delta$ is the local overdensity, and $T_S$ is the spin temperature of the 21\,cm hyperfine transition. The 21\,cm field therefore captures the joint imprint of density, thermal, and ionization fluctuations in the IGM \citep{Furlanetto2006}. Our calculations are based on simulated three-dimensional coeval cubes of $\delta T_\mathrm{b}$ constructed during the EoR by \texttt{LIMFAST}. These cubes encode the spatial structure of ionized regions as well as the large-scale modulation from density and spin temperature fluctuations. The sign and amplitude of the 21\,cm correlation with galaxies depend sensitively on which of these contributions dominates at a given epoch: before ionized bubbles become prevalent, galaxies tend to reside in overdense regions associated with enhanced 21\,cm emission, whereas later in reionization they preferentially trace ionized regions and thus the correlation becomes negative. Throughout, we construct the 21\,cm field with the line-of-sight redshift space distortions (RSDs) from peculiar velocities included \citep{Jensen2013}, whereas the galaxy field is constructed without including RSDs. The 21\,cm foreground and photo-$z$ window functions introduced below are therefore applied in redshift space and real space, respectively\footnote{We note that this asymmetric treatment is motivated by the different effective biases of the two fields because the fractional large-scale RSD correction scales as $1/b$. While Roman LBGs at the redshifts of interest are strongly biased, with $b_\mathrm{gal} \sim 6$--12 (see appendix~\ref{sec:uvlf}) that makes the Kaiser correction small (the additional small-scale suppression from random LOS velocities is likewise negligible compared against the photo-$z$ error), the much lower 21\,cm bias produces more significant RSD effects \citep{Majumdar2013,Majumdar2016}. A fully consistent RSD treatment of both fields and the corresponding filtering and covariance calculations is deferred to future work.}.

To construct the statistic studied in this paper, we first apply foreground filtering to the 21\,cm field in Fourier space, as described in section~\ref{sec:models:obs}, removing the long-wavelength LOS modes strongly contaminated by the spectrally smooth foreground emission. Denoting the filtered 21\,cm field by $\tilde{\mathcal{T}}(\mathbf{k}) = \tilde{\delta T_b}(\mathbf{k})W_\mathrm{21}(\mathbf{k})$, we define a squared field $\mathcal{A}$ through the convolution in Fourier space
\begin{equation}
\tilde{\mathcal{A}}(\mathbf{k}) = \int \frac{d^3 q}{(2\pi)^3} \tilde{\mathcal{T}}(\mathbf{k}-\mathbf{q}) \tilde{\mathcal{T}}(\mathbf{q}) = \int \frac{d^3 q}{(2\pi)^3} W_\mathrm{21}(\mathbf{q}) W_\mathrm{21}(\mathbf{k}-\mathbf{q}) \tilde{\delta T_\mathrm{b}}(\mathbf{q}) \tilde{\delta T_\mathrm{b}}(\mathbf{k}-\mathbf{q}).
\end{equation}
Since the filtering of the 21\,cm field is anisotropic, $\mathcal{A}$ depends on both the orientation and magnitude of $\mathbf{k}$. As detailed below, we rely on this squared field to extract the complementary information between photometrically selected LBGs and the foreground-filtered 21\,cm signal that cannot be measured by directly cross-correlating the two fields. While the direct LBG--21\,cm cross-power spectrum measures how $\delta T_\mathrm{b}$ is enhanced or suppressed around galaxies, the cross-correlation involving $\mathcal{A}$ measures how the \emph{variance} of $\delta T_\mathrm{b}$ is modulated. 

\subsection{Photometrically selected galaxies}

Let $n_\mathrm{gal}(\mathbf{x}, z)$ denote the comoving number density of galaxies selected above a survey-dependent limiting magnitude. The corresponding galaxy overdensity field is
\begin{equation}
\delta_\mathrm{gal}(\mathbf{x}, z) = \frac{n_\mathrm{gal}(\mathbf{x}, z)}{\bar{n}_\mathrm{gal}(z)} - 1,
\end{equation}
where $\bar{n}_\mathrm{gal}(z)$ is the mean galaxy number density at redshift $z$. In practice, we define the detected population through a redshift-dependent selection function of the star formation rate (SFR) corresponding to a given limiting magnitude $m_\mathrm{AB,lim}$. For our analysis, we focus on photometric samples of LBGs to be measured by wide area surveys of the Roman Space Telescope. Relative to spectroscopic surveys, selection of high-$z$ galaxy candidates using the photometric dropout technique offers a much more efficient way to perform a census of the source population driving reionization, albeit at the cost of degraded redshift information. Compared with $\sigma_z \sim 0.01$ for slitless spectroscopy, redshift uncertainties can be 10--100 times larger for the broadband photometrically selected samples and additionally susceptible to catastrophic redshift failures, including contamination by interlopers. This tradeoff is central to the challenge that this work aims to address: while the transverse clustering information remains well preserved, photometric redshift uncertainties smooth the galaxy field along the radial direction and limit access to high $k_\parallel$ modes, where most 21\,cm information resides because low-$k_\parallel$ modes that are contaminated by spectrally smooth foregrounds. 

We incorporate this loss of radial information by relating the actual galaxy overdensity field to the intrinsic one through
\begin{equation}
\delta_\mathrm{gal}(\mathbf{k}) = W_\mathrm{gal}(k_\parallel)\,\delta^{\rm int}_\mathrm{gal}(\mathbf{k}),
\end{equation}
where $W_\mathrm{gal}(k_\parallel)$ describes the impact of photometric redshift uncertainty (see section~\ref{sec:models:obs}). The observed galaxy field always contains shot noise associated with the finite number of detected sources. Since photometric surveys are expected to identify large samples of galaxies over wide areas, the shot-noise contribution is lower compared with spectroscopic samples, though it remains relevant on small scales and at the highest redshifts (see figure~\ref{fig:signals}). 

\subsection{Observational effects} \label{sec:models:obs}

For 21\,cm foreground mitigation, we consider a high-pass filter for low-$k_{\parallel}$ LOS modes with $|k_{\parallel}| \leq k_{\parallel,0}$, in addition to which we further zero out modes within the foreground wedge, $|k_{\parallel}| \leq m k_{\perp}$, with the wedge slope
\begin{equation}
m = \frac{r(z) H(z) \sin\theta}{c(1+z)}, 
\end{equation}
where $r(z)$ is the comoving radial distance and $\theta$ is the angular radius of the field of view. The resulting foreground-filtering window function that vanishes if $|k_{\parallel}| \leq \max(k_{\parallel,0}, mk_{\perp})$ is
\begin{equation}
W_{\rm fg}(k_\parallel,k_\perp)
=
\Theta\!\left(|k_\parallel| - k_{\parallel,0}\right)
\Theta\!\left(|k_\parallel| - m k_\perp\right),
\label{eq:foreground_mask}
\end{equation}
where $\Theta(x) = 1$ if $x>0$ and 0 otherwise. 

On the other hand, the redshift error of galaxies defines a window function that damps the galaxy clustering power
\begin{equation}
W_\mathrm{gal}(k_{\parallel}, z) = \exp \left[ -\frac{1}{2} k^2_{\parallel} \sigma^2_r \right], 
\end{equation}
where $\sigma_{r} = c \sigma_z / H(z)$. Given the considerable uncertainty in 21\,cm foreground contamination and photometric redshift errors, we adopt fixed values of $k_{\parallel,0}$, $m$, and $\sigma_z$ without keeping track of their detailed redshift dependence. In our fiducial case, we assume $\sigma_z = 0.1$, $k_{\parallel,0} = 0.1\,h\mathrm{Mpc^{-1}}$, and $m=1$. 

\begin{figure*}[!ht]
 \centering
 \includegraphics[width=0.85\textwidth]{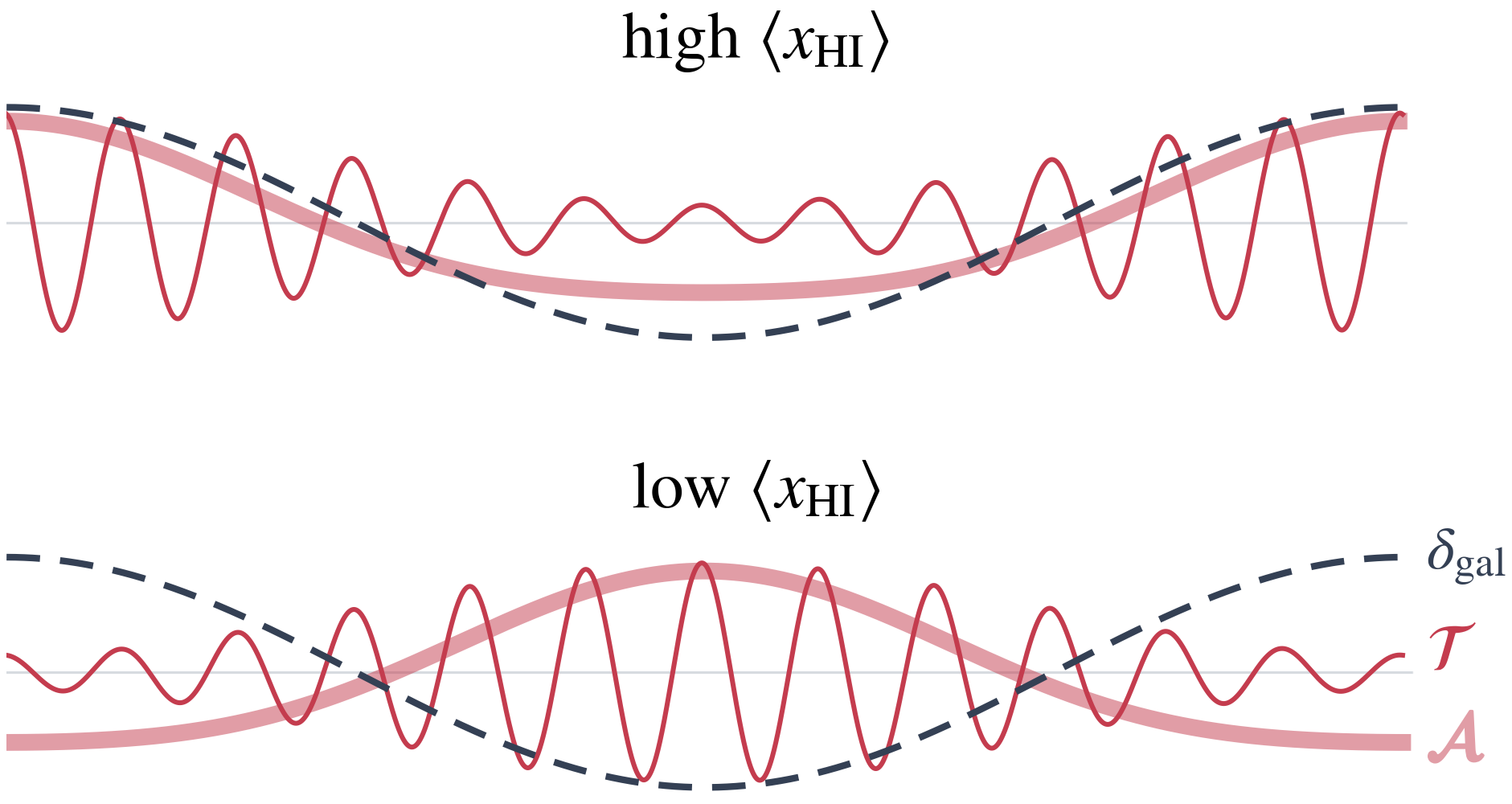}
 \caption{A schematic illustration of a long-wavelength galaxy overdensity field $\delta_{\rm gal}$, the foreground-filtered 21\,cm field ($\mathcal{T}$), and its square ($\mathcal{A}\equiv\mathcal{T}^2$). An individual surviving short-wavelength mode of $\mathcal{T}$ (thin red curve) does not directly form a two-point correlation with the long-wavelength $\delta_{\rm gal}$ mode (gray dashed curve). Squaring $\mathcal{T}$, however, couples pairs of short-wavelength modes to produce a long-wavelength mode in $\mathcal{A}$ (thick red curve), allowing the local small-scale 21\,cm variance to correlate with $\delta_{\rm gal}$. This two-point correlation between $\mathcal{A}$ and $\delta_\mathrm{gal}$, in the $T_{S} \gg T_\mathrm{CMB}$ limit, is positive during the early stages of reionization and negative during the late stages. See section~\ref{sec:results} for more details.}
 \label{fig:cartoon}
\end{figure*}

For a direct cross-correlation between photometric galaxies and the foreground filtered 21\,cm field, the cross-power spectrum can be written as
\begin{equation}
\langle \tilde{\delta}_\mathrm{gal}(\mathbf{k}) \tilde{\mathcal{T}}^{*}(\mathbf{k}') \rangle = \langle \tilde{\delta}^\mathrm{int}_\mathrm{gal}(\mathbf{k}) \delta \tilde{T}^{*}_\mathrm{b}(\mathbf{k}') \rangle W_\mathrm{gal}(k_{\parallel}) W_\mathrm{21}(\mathbf{k}') = (2\pi)^3 \delta^{(3)}_D(\mathbf{k}-\mathbf{k}') P_{\mathrm{gal} \times \mathcal{T}}(\mathbf{k}),
\end{equation}
which becomes strongly suppressed once the product of $\sigma_r$ and $\max(k_{\parallel,0}, m k_{\perp})$ exceeds unity, making the measurement of the cross-correlation extremely challenging even if the signal does not statistically vanish. By contrast, for a cross-correlation with the filter-then-squared field, $\mathcal{A}$, the cross-power spectrum becomes
\begin{equation}
\langle \tilde{\delta}_\mathrm{gal}(\mathbf{k}) \tilde{\mathcal{A}}^{*}(\mathbf{k}') \rangle = (2\pi)^3 \delta^{(3)}_D(\mathbf{k}-\mathbf{k}') P_{\mathrm{gal} \times \mathcal{A}}(\mathbf{k}),
\end{equation}
where
\begin{equation}
P_{\mathrm{gal} \times \mathcal{A}}(\mathbf{k}) = W_\mathrm{gal}(k_{\parallel}) \int \frac{d^3 q}{(2\pi)^3} W_\mathrm{21}(\mathbf{q}) W_\mathrm{21}(\mathbf{k}-\mathbf{q}) B_{\mathrm{g}TT}(-\mathbf{k}, \mathbf{q}, \mathbf{k}-\mathbf{q}) 
\label{eq:PgalA}
\end{equation}
with
\begin{equation}
\langle \tilde{\delta}^\mathrm{int}_\mathrm{gal}(-\mathbf{k'}) \delta \tilde{T}_\mathrm{b}(\mathbf{q}) \delta \tilde{T}_\mathrm{b}(\mathbf{k}-\mathbf{q}) \rangle = (2\pi)^3 \delta^{(3)}_D(\mathbf{k}-\mathbf{k}') B_{\mathrm{g}TT}(-\mathbf{k'}, \mathbf{q}, \mathbf{k}-\mathbf{q}). 
\end{equation}
Thus, $P_{\mathrm{gal} \times \mathcal{A}}(\mathbf{k})$ need not suffer the same mode overlap suppression thanks to mode coupling induced by the higher-order correlation, although its amplitude still depends on the available triangle configurations and the underlying cross-bispectrum $B_{\mathrm{g}TT}$. A schematic illustration of this mechanism is shown in figure~\ref{fig:cartoon}. 

\subsection{Detectability of the LBG--21\,cm$^2$ cross-correlation} \label{sec:models:detectability}

To estimate the thermal noise for SKA-Low, we use
\texttt{21cmSense} v2.2.1 \citep{Murray2024} and consider the full 512-station AA4 design baseline configuration. We adopt a coeval bandwidth of
$\mathcal{B}=15\,\mathrm{MHz}$ sampled by 100 frequency channels and
include baselines shorter than 1000\,m. A total integration time of 1000\,hours is divided equally between two pointings, corresponding to 500\,hours per pointing, which together cover the $\sim10\,\mathrm{deg}^2$ area overlapping the Deep-Tier Roman High-Latitude Wide-Area Survey (HLWAS; \citep{RomanSurveys2025}). Our calculation adopts the default \texttt{21cmSense} sky and receiver
temperature models for the system temperature $T_{\rm sys}(\nu)=T_{\rm sky}(\nu)+T_{\rm rcv}$. The baseline-dependent effective integration times and resulting
nonuniform thermal noise power are evaluated using the actual AA4
station distribution rather than a constant baseline density. The effective noise temperature for each
sampled $\mathbf{u}$ is
\begin{equation}
 T_{\rm rms}(\mathbf{u},z) = \frac{T_{\rm sys}[\nu(z)]}
 {\sqrt{2 b \,t_{\rm eff}(\mathbf{u})}},
\end{equation}
where $\nu(z)=\nu_{21}/(1+z)$ and $t_{\rm eff}(\mathbf{u})$ is set by the AA4 baseline
sampling and observing strategy. The resulting thermal noise
power is
\begin{equation}
 P_\mathrm{n}(\mathbf{u},z) =
 X^2(z) Y(z) \Omega_{\rm eff}(z) \mathcal{B} 
 T_{\rm rms}^2(\mathbf{u},z) =
 X^2(z)Y(z) \Omega_{\rm eff}(z)
 \frac{T_{\rm sys}^2[\nu(z)]}
 {2t_{\rm eff}(\mathbf{u})},
\end{equation}
where $X$ and $Y$ convert angular and frequency intervals to
transverse and radial comoving distances, respectively, and $\Omega_{\rm eff}$ is the effective beam area. We map each baseline cell to the transverse cosmological
wavevector according to
$\mathbf{k_\perp}=2\pi|\mathbf{u}|/X$. This produces a nonuniform
$P_\mathrm{n}(\mathbf{k}_\perp,z)$ that retains the AA4 Fourier-space sampling and is extended over the accessible $k_\parallel$ modes set by the frequency channels. Modes outside the array and channel support are not included. The filtered thermal noise power is then
\begin{equation}
 P_{\mathrm{n},\mathcal{T}}(\mathbf{k})
 =
 \left|W_{21}(\mathbf{k})\right|^2 P_\mathrm{n}(\mathbf{k}).
\end{equation}
where the effective 21\,cm filter is
\begin{equation}
 W_{21}(\mathbf{k}) =
 W_{\mathrm{fg}}(\mathbf{k})
 W_\mathrm{n}(\mathbf{k}),
\end{equation}
where we use the spherically averaged 21\,cm signal power spectrum measured from the fiducial LIMFAST simulation, $P_T^{\mathrm{fid}}$, to define a Wiener filter
\begin{equation}
W_\mathrm{n}(\mathbf{k})
=
\frac{P_T^{\rm fid}(\mathbf{k})}
     {P_T^{\rm fid}(\mathbf{k})+P_\mathrm{n}(\mathbf{k})}, 
\label{eq:wiener_filter}
\end{equation}
which down-weights the noisy modes, preventing them from spreading and dominating in the squared field relevant to our cross-correlation statistic. 

Squaring the filtered 21\,cm field creates both signal--noise and noise--noise contributions to the net covariance. Denoting the filtered cosmological 21\,cm field by $\mathcal{T}$ and its filtered thermal noise by $\mathcal{N}$, the observed squared field contains $\mathcal{T}^2(\equiv\mathcal{A})$, $2\mathcal{T}\mathcal{N}$, and $\mathcal{N}^2$. Assuming Gaussian
thermal noise that is uncorrelated with the cosmological signal and galaxy field, the total noise-induced contribution to the auto-power of $\mathcal{A}$ is
\begin{equation}
 P_{\mathrm{n},\mathcal{A}}(\mathbf{k})
 =
 P_{2\mathcal{T}\mathcal{N}}(\mathbf{k})
 +
 P_{\mathcal{N}^2}(\mathbf{k})
 =
 2\int\frac{d^3q}{(2\pi)^3}
 \left[ 2 P_{\mathcal T}(\mathbf{q})
 +
 P_{\mathrm{n},\mathcal{T}}(\mathbf{q}) \right]
 P_{\mathrm{n},\mathcal{T}}(\mathbf{k}-\mathbf{q}). 
 \label{eq:quadratic_noise_total}
\end{equation}
Here $P_{\mathcal T}=|W_{21}|^2P_T$ is the filtered
21\,cm auto power. The factors of 4 and 2 follow from the Wick contractions of the $2\mathcal{T}\mathcal{N}$ and $\mathcal{N}^2$ terms, respectively. 

Following \citep{FurlanettoLidz2007,Lidz2009}, we can express the variance (in the disconnected Gaussian approximation) of the galaxy--21\,cm$^2$ cross power as
\begin{equation}
\left[ \delta P_{\mathrm{gal} \times \mathcal{A}} (k, \mu) \right]^2 = \frac{1}{2} \left[ P^2_{\mathrm{gal} \times \mathcal{A}} (k, \mu) + \delta P_\mathrm{gal}(k, \mu) \delta P_\mathcal{A}(k, \mu) \right], 
\label{eq:varcross}
\end{equation}
where
\begin{equation}
\delta P_\mathrm{gal}(k, \mu) = P_\mathrm{gal}(k, \mu) + n^{-1}_\mathrm{gal}
\end{equation}
and
\begin{equation}
\delta P_\mathcal{A}(k, \mu) = P_\mathcal{A}(k, \mu) + P_{\mathrm{n},\mathcal{A}}(k, \mu). 
\end{equation}
We note that more generally the full covariance between two binned cross-power estimates can be written schematically as
$\mathrm{Cov}[P_{\mathrm{gal},\mathcal{A}}(\mathbf{k}),P_{\mathrm{gal},\mathcal{A}}(\mathbf{k'})] = \delta^{\rm K}_{\mathbf{k}\mathbf{k'}} (\delta P_{\mathrm{gal} \times \mathcal{A}})^2 + \mathscr{T}^{\delta_{\rm gal} \mathcal{A} \delta_{\rm gal} \mathcal{A}}(\mathbf{k},\mathbf{k'}) / V$,
where $\delta^{\rm K}$ is the Kronecker delta and $V$ is the survey volume. The first term is the disconnected contribution constructed from products of two-point functions, whose diagonal per-mode component is given by eq.~(\ref{eq:varcross}), and the second term scales with the connected cross-trispectrum for the four-point correlation $\langle \delta_{\rm gal} \mathcal{A} \delta_{\rm gal} \mathcal{A} \rangle$. In what follows, we base our sensitivity analysis only on the former. Our uncertainty estimate therefore includes sample variance, galaxy shot noise, and 21\,cm thermal noise, but neglects connected higher-order covariance. The omitted cross-trispectrum term can modify the diagonal variance and induce correlations between different $\mathbf{k}$ modes. This omission may lead to underestimated uncertainties, particularly on small scales and late in reionization, because $\mathcal{A}$ is non-Gaussian even if $\mathcal{T}$ were Gaussian, while patchy reionization can make $\mathcal{T}$ itself substantially non-Gaussian \citep{Mondal2016,Shaw2020}. Its relative impact on the total forecast errors may nevertheless be reduced where thermal noise dominates. 

Letting $N_{\mathrm{m},i}$ be the number of independent Fourier modes of $\mathcal{A}$ in the $i$th $\mu$ bin, we determine and use for analysis those accessible given the binary foreground filtering of the 21\,cm field and the assumed survey specifications. The continuous filters $W_\mathrm{n}$ and $W_\mathrm{gal}$ enter the cross-power spectra and their covariance, but not the mode count. For visualization, we plot the following one-dimensional cross-power spectrum weighted by the mode count,
\begin{equation}
\hat{P}_{\rm gal\times \mathcal{A}}(k) = 
\frac{\displaystyle\sum
N_{\mathrm{m},i}(k)
P_{\rm gal \times \mathcal{A}}(k,\mu_i)}
{\displaystyle\sum N_{\mathrm{m},i}(k)}.
\label{eq:count_weight_P}
\end{equation}
Defining the variance of each binned measurement as $C_i(k) = [\delta P_{\mathrm{gal}\times\mathcal{A}}(k, \mu_i)]^2 / N_{\mathrm{m},i}(k)$ following eq.~(\ref{eq:varcross}), we have
\begin{equation}
[\delta \hat{P}_{\rm gal \times \mathcal{A}}(k)]^2 = \frac{\displaystyle\sum
N^2_{\mathrm{m},i}(k)
C_i(k)}
{\displaystyle\left[\sum N_{\mathrm{m},i}(k) \right]^2} =
\frac{\displaystyle\sum
N_{\mathrm{m},i}(k)
[\delta P_{\rm gal \times \mathcal{A}}(k,\mu_i)]^2}
{\displaystyle\left[\sum N_{\mathrm{m},i}(k) \right]^2}.
\label{eq:count_weight_varP}
\end{equation}
The cumulative detection significance, obtained by inverse-variance weighting cross-power spectra measured in $(k, \mu)$, can be calculated as
\begin{equation}
\left(\frac{S}{N}\right)^2
=
\sum_{j} \sum_{i} \frac{P_{\rm gal\times \mathcal{A}}^2(k_j,\mu_i)}{C_i(k_j)}
=
\sum_{j} \sum_{i}
\frac{N_{\mathrm{m},i}(k_j) P_{\rm gal\times \mathcal{A}}^2(k_j,\mu_i)}
{[\delta P_{\rm gal\times \mathcal{A}}(k_j, \mu_i)]^2}.
\label{eq:snr}
\end{equation}
Note that eq.~(\ref{eq:count_weight_P}) and (\ref{eq:count_weight_varP}) describe the cross-power spectra and their uncertainties shown in, e.g., figure~\ref{fig:constraints}, whereas eq.~(\ref{eq:snr}) is used to evaluate the S/N quoted using the resolved anisotropic $P_{\mathrm{gal}\times\mathcal{A}}(k,\mu)$ measurements, providing optimal weighting under the adopted covariance approximation. The Fisher forecasts described below are based on these same resolved measurements and their covariance. With this formalism in place, we focus on the detectability and the constraining power on EoR parameters of the cross-power spectrum in what follows. For completeness, we present the corresponding detectability formalism for the cross-correlation coefficient in appendix~\ref{sec:ccc}.


\subsection{Fisher matrix analysis of the astrophysical parameters} \label{sec:models:fisher}

To forecast the constraints on key EoR parameters obtainable from the LBG--21\,cm$^2$ cross-correlation, we utilize the Fisher information matrix, whose inverse defines the theoretical lower limit on the parameter uncertainties for an unbiased estimator (the Cramér-Rao bound). Specifically, assuming Gaussian variance, we define the Fisher matrix for the EoR parameter vector $\boldsymbol{\theta}= \{ f_\mathrm{esc}, L_X, \alpha_{\eta}, T_\mathrm{vir,min} \}$ as
\begin{equation}
F_{ij} = \sum_{z,k,\mu} \frac{1}{\mathrm{var}(P_\mathrm{gal\times\mathcal{A}})} \frac{\partial P_\mathrm{gal\times\mathcal{A}}}{\partial \theta_i} \frac{\partial P_\mathrm{gal\times\mathcal{A}}}{\partial \theta_j}. 
\end{equation}
To ensure numerical convergence (which we explicitly verify), we calculate these derivatives using central differences based on 8 additional sets of \texttt{LIMFAST} simulations generated with 5--10\% offsets from the fiducial parameter values.

\section{Simulations} \label{sec:sims}

\subsection{The \texttt{LIMFAST} framework and its astrophysical models}

To generate the mock 21\,cm and galaxy fields outlined in section~\ref{sec:models}, we use \texttt{LIMFAST}, a semi-numerical simulation framework built on the \texttt{21cmFAST} code \citep{Mesinger2011,Park2019,Murray2020} and extended to more general applications involving cosmic background radiation, especially line intensity mapping (LIM) signals, during cosmic dawn and the EoR. Starting from cosmological initial conditions, \texttt{LIMFAST} evolves the matter density field using second-order Lagrangian perturbation theory and constructs the structure formation and ionization history using the excursion set approach. \texttt{LIMFAST} tracks halo specific quantities relevant for source modeling, such as the SFR, gas mass, metallicity, and multiwavelength spectral emissivities, derived from a subgrid analytic galaxy formation model \citep{Furlanetto2017,Furlanetto2021}, thereby enabling physically grounded predictions for both the 21\,cm signal and other tracers of large-scale structure. Specifically, the general code architecture and methodology are introduced in Papers I and II \citep{MasRibas2023,Sun2023}, whereas Papers III and IV demonstrate its application to higher-order reionization statistics \citep{Sun2025} and simulation-based inference of galaxy formation physics \citep{Sun2026}.

Ionized regions are identified using the excursion set formalism \citep{Bond1991,Furlanetto2004}, which accounts for both the local production of ionizing photons by star formation and inhomogeneous recombinations \citep{SobacchiMesinger2014}. Meanwhile, Ly$\alpha$ coupling through the Wouthuysen--Field effect and X-ray heating of the IGM together determine the 21\,cm spin temperature field. The four astrophysical parameters, later varied in our Fisher analysis (section~\ref{sec:results:detect}), control different parts of this calculation. The parameter $\alpha_\eta$ sets the halo-mass dependence of the stellar-feedback mass loading and hence the suppression of star formation in low-mass halos; $f_{\rm esc}$ specifies the fraction of ionizing photons that escape into the IGM; $L_X/{\rm SFR}$ sets the X-ray luminosity per unit star formation; and $T_{\rm vir,min}$ determines the minimum virial temperature of star-forming halos. Among others, these parameters control the source population and the resulting ionization and thermal histories of the IGM.

Taking into account dust attenuation empirically determined from HST and JWST data \citep{ZhaoFurlanetto2024}, we calibrate our subgrid galaxy model such that observational constraints on galaxy UV luminosity functions (UVLFs) at $5<z<10$ are reasonably reproduced. Appendix~\ref{sec:uvlf} presents the corresponding comparison with HST and JWST measurements and provides details about the dust treatment. The predicted clustering bias of Roman-detectable LBGs considered in this work is also presented. 

\subsection{Hybrid halo field treatment and galaxy selection}

\texttt{LIMFAST} constructs mock galaxy fields tied to the same underlying source population for the EoR and the 21\,cm signal, making it well suited for cross-correlation studies where the galaxy and IGM observables must be modeled within a physically connected framework. When external initial conditions and halo catalogs from high-resolution $N$-body simulations are supplied, \texttt{LIMFAST} can base its calculations directly on the resulting discrete halo populations, thereby improving the realism of the halo evolution, abundance, and spatial clustering, all of which are essential for reliably modeling and interpreting the 21\,cm--galaxy correlation of interest. In this work, we adopt the initial conditions and halo catalogs at $z \gtrsim 6$ from the mini-Uchuu suite of the Uchuu simulations \citep{Ishiyama2021} in a (512\,cMpc)$^3$ volume sampled on a 256$^3$ grid. Given the mass resolution of mini-Uchuu, $m_\mathrm{part}=3.3\times10^8\,M_{\odot}/h$, only halo samples above $\sim 10^{10}\,M_{\odot}$ are reliably resolved and mass-complete. To account for the entire ionizing source population needed for the excursion set calculation of reionization, we therefore combine externally resolved halo fields with integrals over unresolved halos described by the conditional halo mass functions \citep{SMT2001} derived from the same initial conditions. This hybrid treatment of resolved and unresolved populations is implemented consistently in the construction of the IGM ionization field, such that bright galaxies hosted by discrete, resolved halos are represented explicitly, whereas the cumulative contribution from lower-mass halos below mini-Uchuu's completeness limit (and above $T_\mathrm{vir,min}\approx10^4\,$K for efficient atomic cooling) is retained and represented statistically. 

For the selection of Roman galaxy samples, the detectable LBGs are drawn from the resolved mini-Uchuu catalog. Each resolved halo is assigned a star formation rate and dust-attenuated UV luminosity using the same model that determines its ionizing output. Galaxies below the mini-Uchuu completeness threshold are not included in the Roman galaxy counts, though their aggregate, density-dependent contribution to reionization is retained. Our hybrid approach therefore captures the environments of the strongly clustered bright LBGs observable by Roman, while preserving the galaxy--halo connection and the impact of unresolved faint sources on the reionization process as prescribed by our underlying feedback-regulated galaxy formation model. All 21\,cm and galaxy fields analyzed are generated from coeval boxes of volume $512$\,cMpc$^3$ (with cell size 2\,cMpc), sufficient for capturing the pertinent survey volumes for our cross-correlation analysis. Consequently, their evolution across the observing bandwidth, namely the light cone effect, is neglected. This allows useful baseline forecasts as presented in this work, though the light cone effect on summary statistics especially during periods of rapid IGM ionization or thermal evolution can be non-trivial \citep{Datta2012,Datta2014,Ghara2015}.

We summarize in table~\ref{tab:assumptions} the key modeling, simulation, and analysis assumptions made throughout this work.

\begin{table}[t]
\centering
\caption{Summary of the modeling, simulation, and analysis
assumptions adopted in this work.}
\label{tab:assumptions}
\scriptsize
\renewcommand{\arraystretch}{1.12}
\begin{tabularx}{\textwidth}{@{}l l X@{}}
\toprule
Description & Reference & Assumptions \\
\midrule

\multicolumn{3}{@{}l}{\itshape Modeling and simulation setup} \\

Cosmology &
Section \ref{sec:intro} &
Planck 2016 cosmology \\

Simulation framework &
Section \ref{sec:sims} &
\texttt{LIMFAST} with embedded mini-Uchuu initial conditions and halo catalogs \\

Redshift sampling &
Section \ref{sec:results} &
$z= 5.72$, 6.34, 7.03, 7.77, 8.59, 9.48, and 10.45 \\

Simulation volume &
Section \ref{sec:results} &
$(512\,{\rm cMpc})^3$ coeval boxes, each sampled on a $256^3$ grid \\

Halo treatment &
Section \ref{sec:sims} &
Resolved mini-Uchuu catalogs above completeness level $M_\mathrm{h} \sim 10^{10}\,M_\odot$, below which
conditional Sheth--Tormen halo mass functions used \\

EoR parameters &
Section \ref{sec:results} &
$f_{\rm esc}=0.1$,
$\log_{10}(L_X/{\rm SFR})=40.5$,
$\alpha_\eta=0.67$, and
$\log_{10}(T_{\rm vir,min}/{\rm K})=4$ \\

Galaxy formation model &
Sections \ref{sec:results}, \ref{sec:uvlf} &
Calibrated to observed UVLFs with redshift-dependent dust attenuation \\

\midrule
\multicolumn{3}{@{}l}{\itshape Roman LBG survey} \\

Galaxy selection &
Section \ref{sec:results} &
LBGs with $m_{\rm AB,lim}=27$ \\

Survey area &
Section \ref{sec:results} &
$10\,{\rm deg}^2$ \\

Photo-$z$ error &
Sections \ref{sec:models:obs}, \ref{sec:results} &
Gaussian error with $\sigma_z=0.1$ (fiducial) and 0.3 (comparison) \\

\midrule
\multicolumn{3}{@{}l}{\itshape SKA-Low observations} \\

Instrument specification &
Section \ref{sec:models:detectability} &
512-station AA4 configuration calculated using
\texttt{21cmSense} v2.2.1 \\

Observing strategy &
Section \ref{sec:models:detectability} &
1000\,hours divided between two pointings, covering $10\,{\rm deg}^2$ \\

Frequency sampling &
Section \ref{sec:results} &
$\mathcal{B} = 15$\,MHz coeval bandwidth with 100 channels \\

Baseline selection &
Section \ref{sec:models:detectability} &
Baselines shorter than $1000$ m \\

Foreground mask &
Section \ref{sec:models:obs} &
$W_{\rm fg}$ defined by $k_{\parallel,0}=0.1\,h\,{\rm Mpc}^{-1}$ and wedge slope $m=1$ \\

Noise weighting &
Section \ref{sec:models:detectability} &
$W_{21}=W_{\rm fg}W_\mathrm{n}$, with a Wiener filter, $W_\mathrm{n}$, included \\

\midrule
\multicolumn{3}{@{}l}{\itshape Sensitivity analysis} \\

Detection significance &
Sections \ref{sec:models:detectability}, \ref{sec:results} &
Mean power spectra with cumulative S/N evaluated
over $k<0.3\,h\,{\rm Mpc}^{-1}$ \\

Fisher forecasts &
Sections \ref{sec:models:fisher}, \ref{sec:results} &
Central derivatives from 8 extra simulations with 5--10\% parameter offsets \\

\bottomrule
\end{tabularx}
\end{table}

\section{Results} \label{sec:results}

\subsection{21\,cm and galaxy overdensity fields}

\begin{figure*}[!ht]
 \centering
 \includegraphics[width=\textwidth]{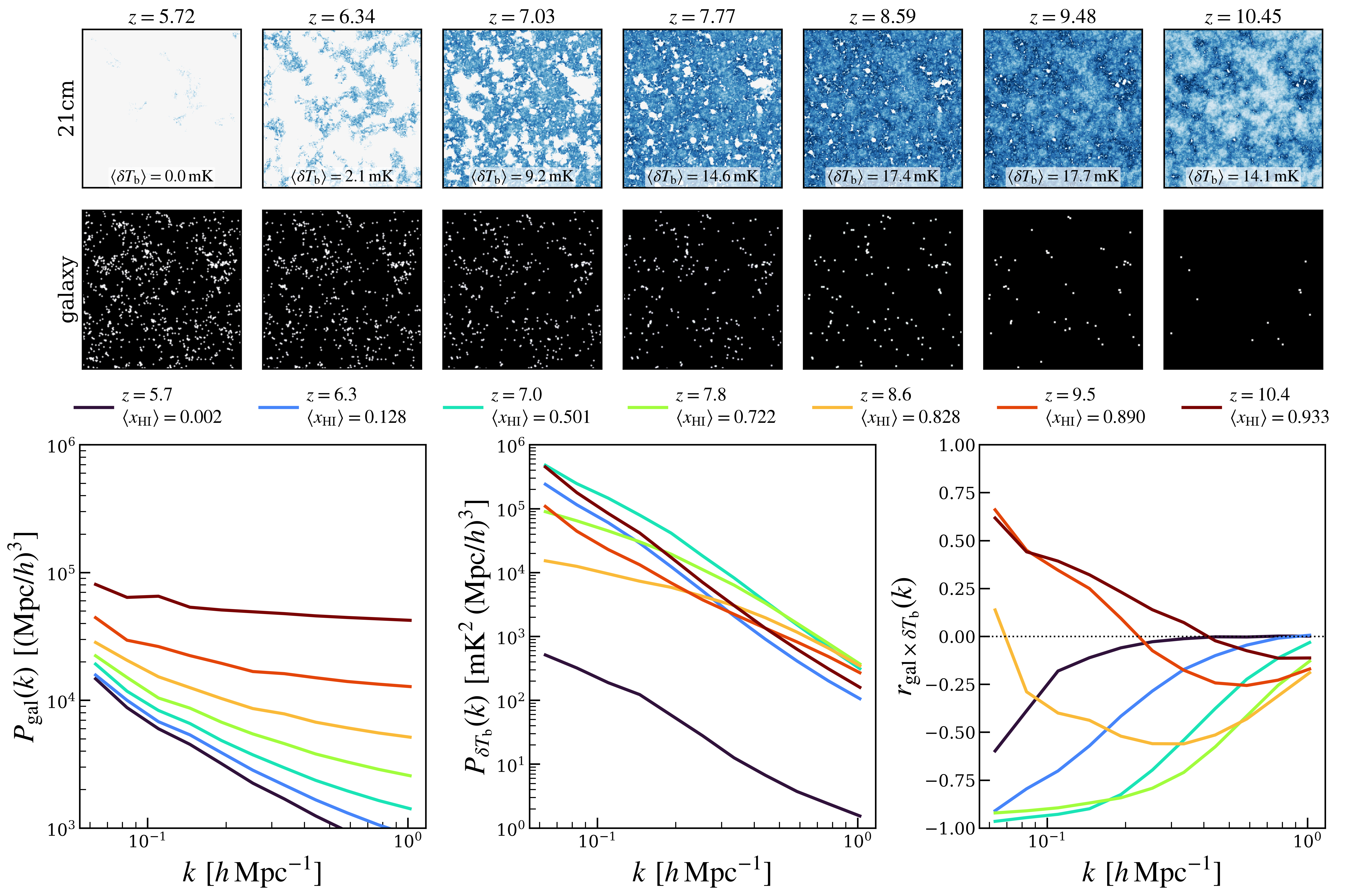}
 \caption{Evolution of the 21\,cm and galaxy overdensity fields during the EoR predicted by our fiducial simulation. The top panels illustrate the opposite IGM phases traced by the 21\,cm brightness temperature and the galaxy distribution, with the former appearing in emission through most of the EoR owing to early IGM heating in our fiducial model. The three bottom panels show the corresponding power spectra of the galaxy overdensity field for a limiting magnitude of $m_\mathrm{AB,lim}=27$ (left), the 21\,cm brightness temperature (middle), and their direct, unfiltered cross-correlation coefficient (right). At very high redshift, the correlation is weaker partly because the galaxy power spectrum is more strongly affected by the shot noise. As reionization proceeds, the 21\,cm--galaxy correlation reverses sign, transitioning from positive at high $\langle x_\mathrm{HI} \rangle$ to negative at low $\langle x_\mathrm{HI} \rangle$.}
 \label{fig:signals}
\end{figure*}

Taking the simulation setup described in section~\ref{sec:sims}, we generate coeval 21\,cm and galaxy fields at $z=5.72$, 6.34, 7.03, 7.77, 8.59, 9.48, and 10.45, corresponding to the redshifts where mini-Uchuu halo catalogs exist. Each simulated box has a side length of 512\,cMpc and is sampled on a $256^3$ grid. The mean IGM neutral fractions at these redshifts are approximately $\langle x_{\rm HI}\rangle=0.002$, 0.13, 0.50, 0.72, 0.83, 0.89, and 0.93, respectively, broadly consistent with the state-of-the-art observational constraints \citep{Robertson2022}. The mock Roman LBG samples are selected using the dust-attenuated UV luminosities predicted by our fiducial galaxy formation model, with an adopted limiting magnitude of $m_{\rm AB,lim}=27$ that represents a conservative choice consistent with the Deep-Tier imaging depth of Roman HLWAS\footnote{The $5\sigma$ point-source limiting magnitudes for Roman's F106, F129, F158, and F184 filters (relevant for dropout selections of EoR galaxies) are 27.7, 27.6, 27.5, and 27.0, respectively.}.

We show in figure~\ref{fig:signals} the simulated evolution of the 21\,cm and galaxy fields. In our fiducial simulation, X-ray heating raises the 21\,cm spin temperature above the CMB temperature before the bulk of reionization, making the 21\,cm signal appear predominantly in emission over the redshift range considered here. At $z\gtrsim10$, the IGM remains mostly neutral and the 21\,cm fluctuations largely follow the matter density, together with spatial variations in the spin temperature and the early ionization field. As redshift decreases, reionization proceeds inside-out, with ionized regions growing mainly around galaxy overdensities first. By $z \lesssim 6$, the remaining 21\,cm signal is concentrated in neutral islands between large ionized regions, before the reionization process completes.

The lower panels of figure~\ref{fig:signals} show the corresponding spherically averaged power spectra. The galaxy power spectrum becomes increasingly dominated by the shot noise toward high redshift owing to the declining number density of galaxies brighter than $m_{\rm AB,lim}=27$. The 21\,cm power spectrum exhibits a non-monotonic evolution determined by the competing density, thermal, and ionization contributions. The direct, unfiltered galaxy--21\,cm cross-correlation is positive on large scales at the highest redshifts when overdense regions host both more galaxies and enhanced 21\,cm emission. It then becomes negative once the ionized regions surrounding galaxies govern the 21\,cm morphology, before dropping to zero at the end of reionization. This sign evolution marks a useful physical reference for our squared statistic studied below, although the sign of the LBG--21\,cm$^2$ cross-correlation does not necessarily follow that of the direct cross-correlation in detail.

\subsection{The non-vanishing LBG--21\,cm$^2$ cross-correlation}

\begin{figure*}[!ht]
 \centering
 \includegraphics[width=\textwidth]{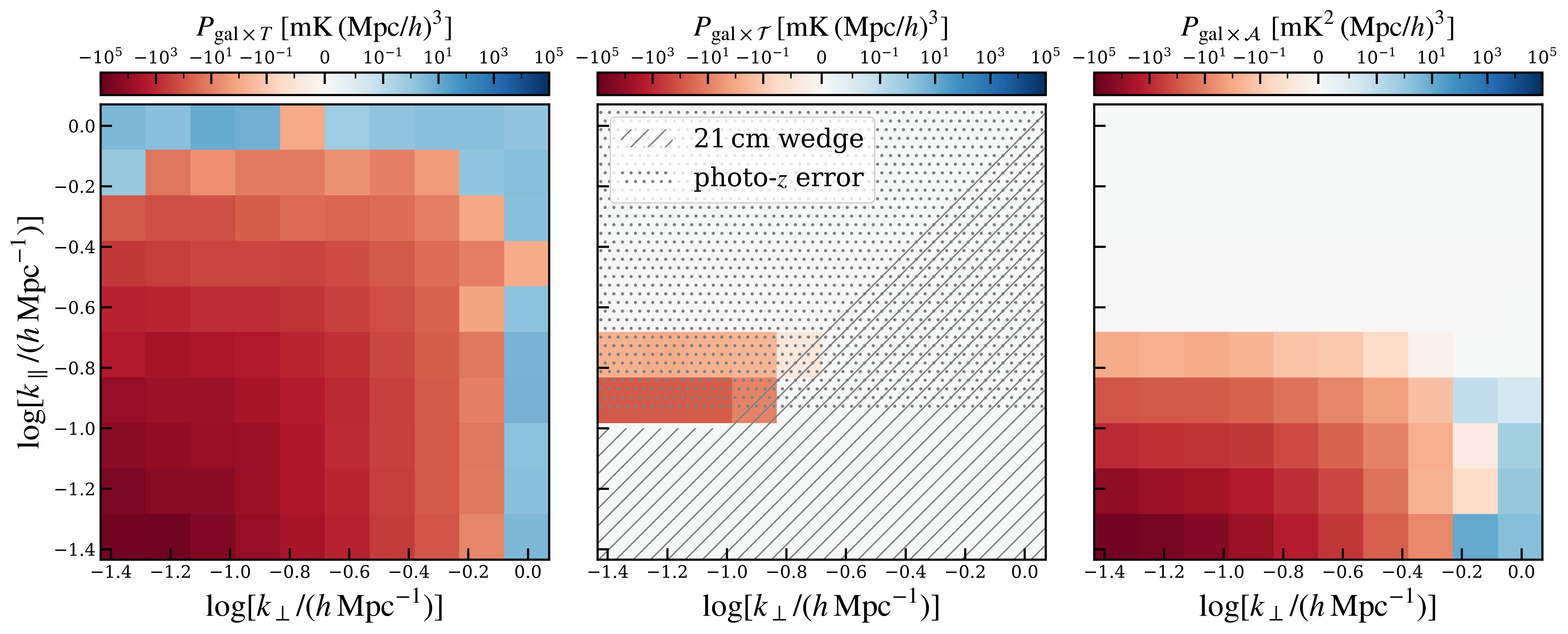}
\caption{Two-dimensional power spectra of the LBG--21\,cm cross-correlation with and without filtering at $z=6.34$, where $\langle x_\mathrm{HI} \rangle \sim 0.13$. \textit{Left:} The direct cross-power spectrum between $\delta^\mathrm{int}_\mathrm{gal}$ and $\delta T_\mathrm{b}$, without filtering by $W_\mathrm{gal}$ or $W_\mathrm{fg}$, shows a strong negative correlation on large scales and turns mildly positive on the smallest scales. \textit{Middle:} The hatched regions illustrate how 21\,cm foreground filtering (including the wedge induced by the chromatic instrument response) together with photometric redshift uncertainty ($\sigma_z=0.1$ assumed here) removes the vast majority of overlap between the Fourier modes accessible to the LBG overdensity and 21\,cm fields filtered by $W_{\rm gal}$ and $W_{\rm fg}$, respectively. The direct cross-correlation signal as a result is effectively undetectable. Note that the continuous Wiener filter, $W_{\rm n}$, is not included here. \textit{Right:} Cross-correlating the galaxy field with the filtered-then-squared 21\,cm field $\mathcal{A}$ produces a non-vanishing signal at low $k_{\parallel}$ through mode coupling, while retaining a similar large-scale negative correlation. The radial extent of the resulting cross-correlation is set by $W_{\rm gal}$, which suppresses modes with $k_\parallel \sigma_r \gtrsim 1$.}
 \label{fig:2dps}
\end{figure*}

\begin{figure*}[!ht]
 \centering
 \includegraphics[width=\textwidth]{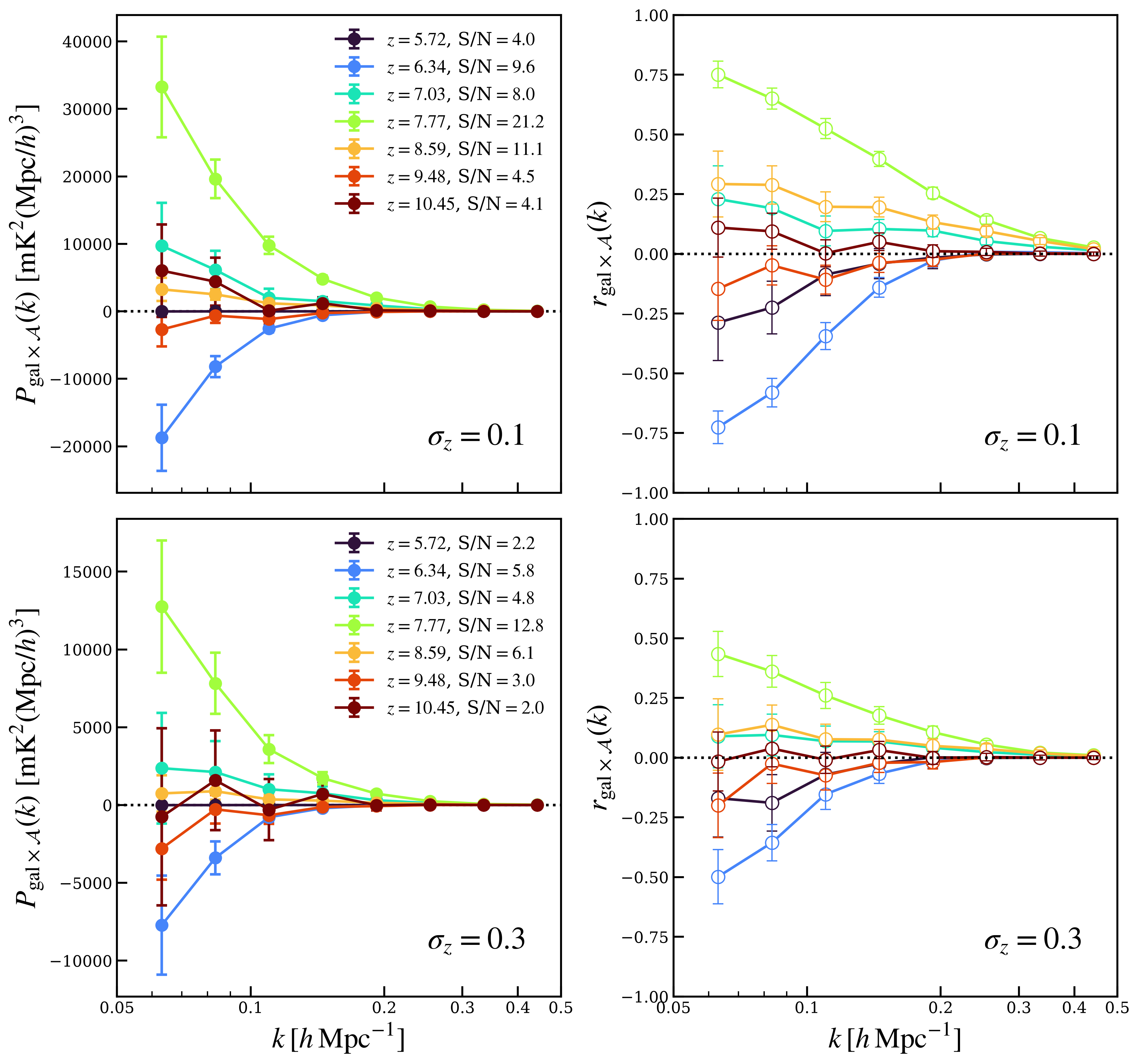}
 \caption{\textit{Top:} The left panel shows the detectability of the cross-power spectrum between Roman-selected LBGs and the filtered-then-squared 21\,cm field measured by SKA-Low, as a function of scale and redshift. We assume an overlapping survey area of $10\,\mathrm{deg}^2$, an optimistic photometric redshift uncertainty of $\sigma_z = 0.1$, and a 21\,cm survey bandwidth of $\Delta\nu = 15\,\mathrm{MHz}$. The quoted signal-to-noise ratios are summed over large-scale bins with $k<0.3\,h/\mathrm{Mpc}$. The right panel shows the cross-correlation coefficient and its detectability (see appendix~\ref{sec:ccc}) for each of the corresponding LBG--21\,cm$^2$ cross-correlation signals. The constraints on $P_{\mathrm{gal}\times\mathcal{A}}(k, \mu)$ are used in Fisher matrix analysis of EoR parameters, whereas those on $r_{\mathrm{gal}\times\mathcal{A}}$ are shown for illustrative purposes only. \textit{Bottom:} The same as the top panels but with a more conservative $\sigma_z = 0.3$.}
 \label{fig:constraints}
\end{figure*}

For the reasons discussed in section~\ref{sec:models:obs}, the direct cross-correlation between foreground-filtered 21\,cm measurements and photometric galaxies is strongly suppressed given that few modes are shared between both tracers. Figure~\ref{fig:2dps} illustrates this effect at $z=6.34$, when the direct, unfiltered 2D cross power spectrum is strongly negative on large scales and becomes less negative and eventually mildly positive toward the smallest scales. This is consistent with the scale dependence of the direct cross-correlation coefficient shown in figure~\ref{fig:signals}. The middle panel shows the same 2D cross power spectrum after imposing the 21\,cm foreground window function, $W_\mathrm{fg}$ and the galaxy window function associated with a photometric redshift uncertainty of $\sigma_z=0.1$. The foreground filter removes the low-$k_\parallel$ modes where LBGs retain most of the clustering information. Even though the direct cross-correlation is not strictly zero after filtering, the Fourier space with appreciable overlap is sufficiently narrow that the signal becomes practically undetectable. As shown in the right panel, this loss of overlap does not impact the filtered-then-squared field in the same way. As the squaring operation convolves pairs of surviving 21\,cm modes, two modes with large, roughly opposite $k_\parallel$ can contribute to a squared-field mode with low net $k_\parallel$ that is accessible to LBGs. Although the individual high-$k_{\parallel}$ 21\,cm modes have little direct overlap with the low-$k_{\parallel}$ galaxy modes, their small-scale variance can be suppressed (or enhanced) around galaxy overdensities. As a result, $P_{{\rm gal}\times\mathcal{A}}$ recovers a clear negative large-scale cross-correlation at this redshift. This mode coupling is the central mechanism through which our higher-order statistic accesses information unavailable to the direct two-point statistic and extends the scope of galaxy--21\,cm synergy during the EoR beyond spectroscopic galaxy samples. 

\subsection{Detectability and constraints on the EoR} \label{sec:results:detect}

\subsubsection{Baseline detectability forecasts}

In figure~\ref{fig:constraints}, we show our SKA-Low detectability forecasts for two contrasting cases: one with an optimistic photometric redshift uncertainty of $\sigma_z = 0.1$ and another with a more conservative value of $\sigma_z = 0.3$, following the sensitivity analysis formalism described in sections~\ref{sec:models:obs} and \ref{sec:models:detectability}. Our forecasts consistently include the SKA-Low AA4 sampling, the foreground mask, the Wiener filter, and the Roman photo-$z$ uncertainty. The error bars include the Gaussian covariance from the cosmological signals, galaxy shot noise, and thermal noise contributions generated when the filtered 21\,cm field is squared. 

\begin{figure*}[!ht]
 \centering
 \includegraphics[width=\textwidth]{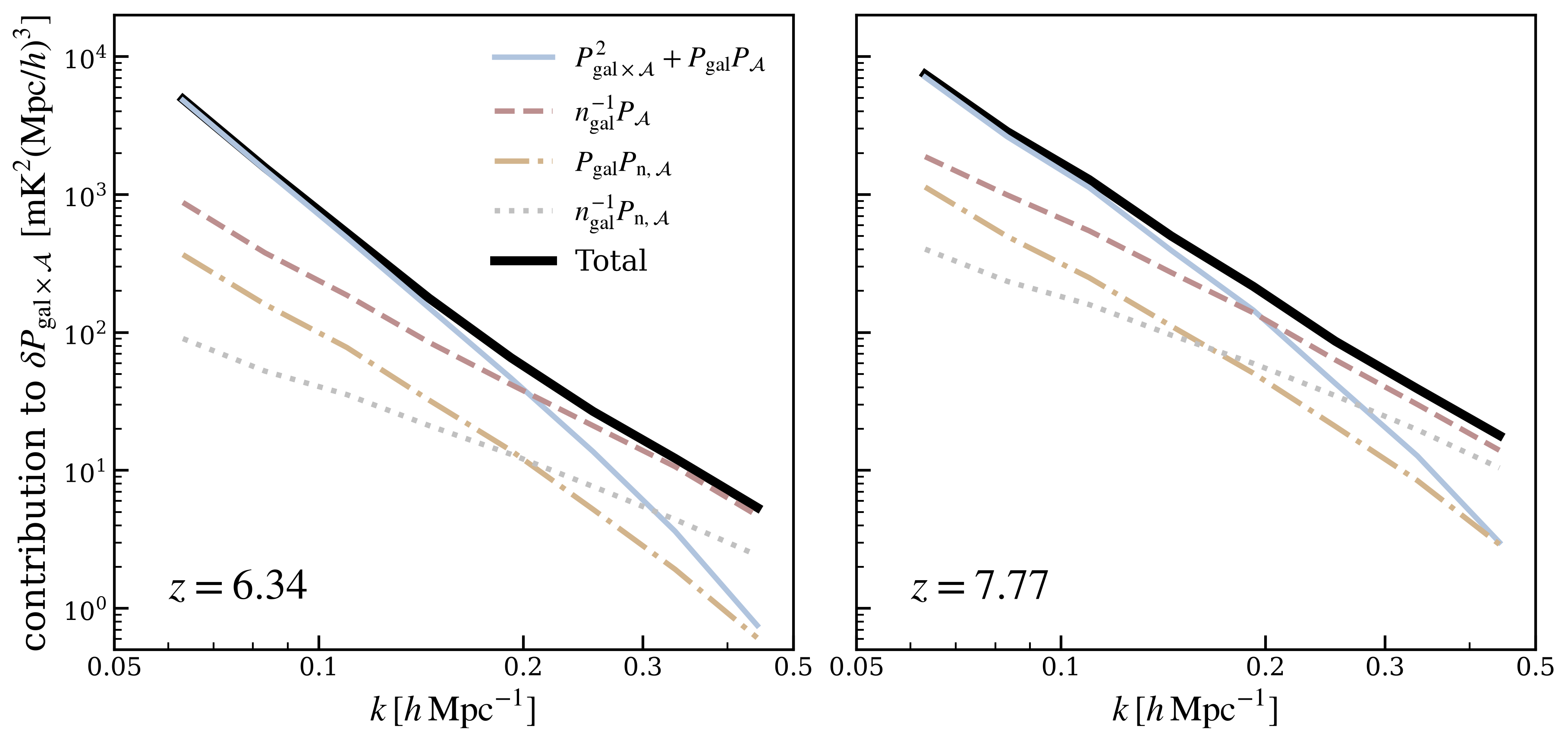}
 \caption{Error budget for the LBG--21\,cm$^2$ cross-power spectrum at $z=6.34$ and 7.77. The square root of the total Gaussian variance shown in black given by eq.~(\ref{eq:varcross}) is decomposed into contributions from sample variance ($P_{\mathrm{gal}\times\mathcal{A}}^2+P_\mathrm{gal} P_{\mathcal{A}}$; blue), galaxy shot noise ($n^{-1}_\mathrm{gal} P_{\mathcal{A}}$; red), 21\,cm thermal noise ($P_\mathrm{gal} P_{\mathrm{n},\mathcal{A}}$; yellow), and their joint contribution ($n^{-1}_\mathrm{gal} P_{\mathrm{n},\mathcal{A}}$; gray). The square root of each contribution is plotted to match the units of $P_{\mathrm{gal}\times\mathcal{A}}$.}
 \label{fig:errbudget}
\end{figure*}

The left panel shows that the LBG--21\,cm$^2$ cross-power spectrum is detectable over a substantial portion of the EoR, although the significance evolves strongly with redshift. Summing over large-scale modes with $k<0.3\,h\,{\rm Mpc}^{-1}$, we find total signal-to-noise ratios ranging from 5 (3) to 20 (13) over $6 \lesssim z \lesssim 10$ for $\sigma_z = 0.1$ (0.3). The strongest detections of the cross-correlation occur at $z=7.77$ and 8.59, when the IGM is mostly neutral yet the ionization field has developed substantial large-scale structure. The strongly negative correlation at $z=6.34$ is also detectable at high significance. By contrast, the detectability gets poor at $z\gtrsim9$, where the cross-correlation is weakened by both the increased galaxy shot noise and more complicated physical contributions (see section~\ref{sec:results:interpret}), and at $z < 6$, where the 21\,cm signal is too weak. The right panel shows that the cross-correlation is strongest on large scales and generally approaches zero toward $k\simeq0.3$--$0.5\,h\,{\rm Mpc}^{-1}$. The large-scale signal is negative at $z=5.72$ and 6.34, becomes positive at $z=7.03$, reaches its maximally positive amplitude at $z=7.77$ and 8.59, and drops to weaker levels at $z=9.48$ and 10.45. The overall evolution of the cross-correlation on large scales is consistent with the evolving reionization environments traced by LBGs. The scale dependence shown reflects in part the smoothing by $W_{\rm gal}$, which confines appreciable cross power to increasingly transverse modes as $k$ increases, along with the effect of galaxy shot noise that grows more important at higher $k$ (see figure~\ref{fig:errbudget}). 

We also assess the covariance approximation in eq.~(\ref{eq:varcross}) used for our forecasts, which neglects the connected cross-trispectrum term. Because the available \texttt{LIMFAST} outputs share a single parent mini-Uchuu realization, we perform a subvolume test where we divide each $(512\,\mathrm{cMpc})^3$ coeval box into 27 subvolumes. We average the covariance matrices obtained from 16 random realizations of the gridded galaxy counts perturbed by $\sigma_z$. We compare these measurements with the average covariance estimated separately from 16 pairs of $\delta_{\rm gal}$ and $\mathcal{A}$ fields---generated to be Gaussian (i.e., no connected four-point correlation) by construction and preserving the same simulated auto- and cross-power spectra. Their difference provides an approximate estimate of the effective connected covariance omitted from eq.~(\ref{eq:varcross}). Over the scales probed by this test ($0.1 \lesssim k \lesssim 0.3\,h\,\mathrm{Mpc}^{-1}$), we find no evidence that the non-Gaussian contribution dominates the covariance. The cumulative S/N changes by no more than $15\%$. Thus, the Gaussian covariance assumed provides a reasonable baseline, although moderate corrections cannot be excluded without an ensemble of independent simulations.

In figure~\ref{fig:errbudget}, we further show how different terms contribute to the predicted uncertainties of the LBG--21\,cm$^2$ cross-power spectrum. On the largest accessible scales, the uncertainties are dominated by sample variance, whereas the relative importance of galaxy shot noise grows toward smaller scales and becomes more dominant than the sample variance contribution starting $k \sim 0.2\,h\,{\rm Mpc}^{-1}$. The terms involving 21\,cm thermal noise remain subdominant over most of the scales relevant for our forecasts, suggesting that for our fiducial survey configuration the measurements are not primarily limited by raw 21\,cm sensitivity. Instead, the finite survey volume governs the detectability on the largest scales, while the limited number of bright LBGs controls the constraints on smaller scales. This implies that simply increasing the integration time of 21\,cm observations alone in this case will not enable substantial gains in the cross-power S/N.

\subsubsection{Dependence on observational assumptions}

\begin{figure*}[!ht]
 \centering
 \includegraphics[width=\textwidth]{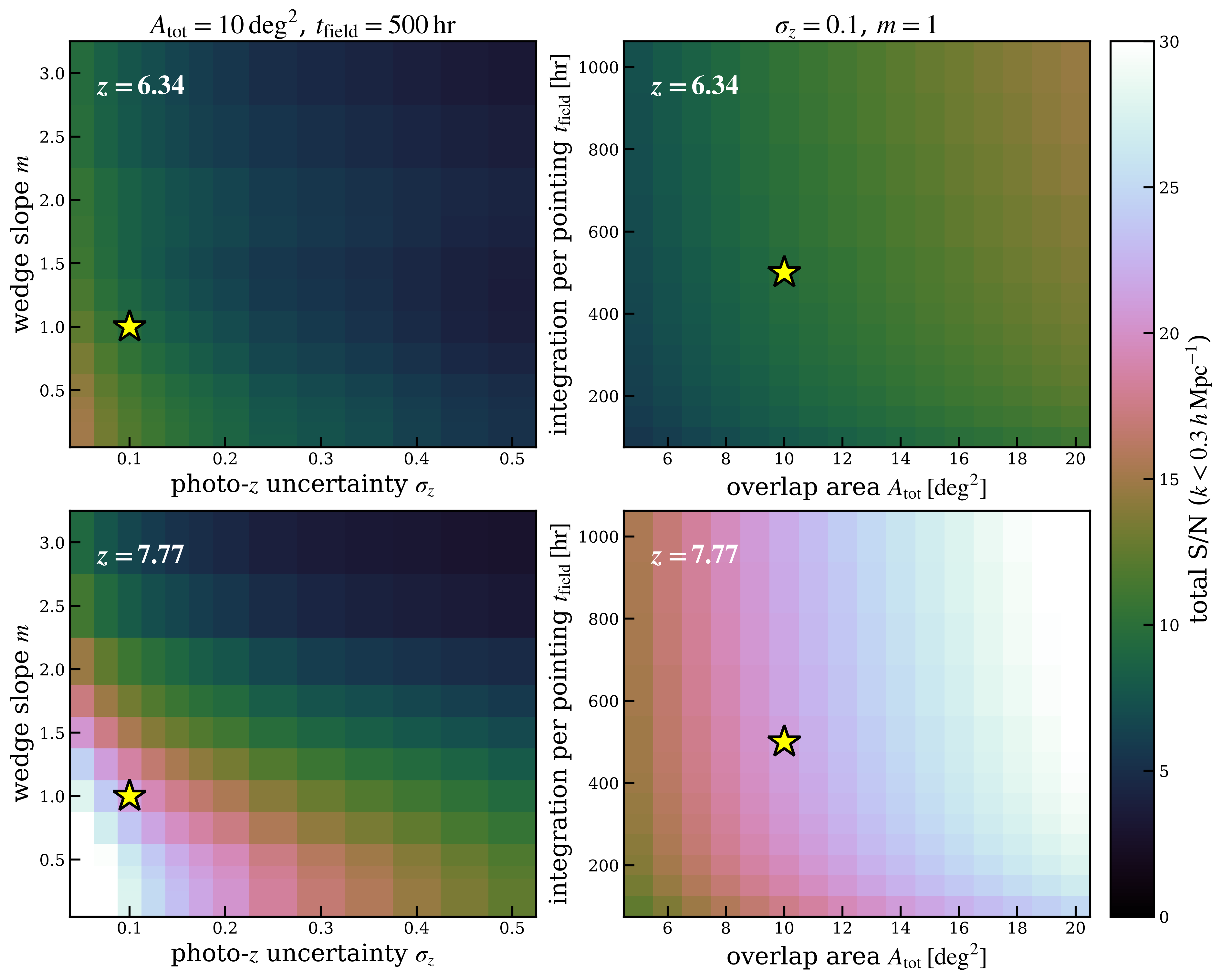}
 \caption{Dependence of the LBG--21\,cm$^2$ cross-power detectability on observational assumptions made at $z=6.34$ (top) and $7.77$ (bottom). The color coding shows the total S/N evaluated for $k<0.3\,h\,{\rm Mpc}^{-1}$. The left column varies the photo-$z$ uncertainty $\sigma_z$ and foreground-wedge slope $m$ while holding the overlapping survey area $A_{\rm tot}=10\,{\rm deg}^2$ and integration time per SKA-Low pointing $t_{\rm field}=500\,{\rm hr}$ fixed. The right column instead varies $A_{\rm tot}$ and $t_{\rm field}$ (without assuming a fixed total SKA-Low observing time) while holding $\sigma_z=0.1$ and $m=1$ fixed. Locations of the yellow stars indicate our fiducial assumptions.}
 \label{fig:survey}
\end{figure*}

Extending the detectability analysis, we illustrate in figure~\ref{fig:survey} how the predicted cross-power S/N depends on the assumptions made about photometric redshift error, 21\,cm foreground cleaning, and the survey configuration. As shown in the left column, increasing either the photo-$z$ error $\sigma_z$ or the foreground wedge slope $m$ deteriorates the detectability, though through different mechanisms. A poorer $\sigma_z$ more strongly damps the low-$k_{\parallel}$ modes accessible by the galaxy field, whereas a steeper $m$ removes a larger fraction of the high-$k_{\parallel}$ 21\,cm modes before the field is squared. Squaring couples pairs of surviving modes with approximately opposite $k_\parallel$ into modes with small net $k_\parallel$, but it cannot recover information carried by modes already removed by filtering. The resulting statistic is therefore resilient to yet dependent on foreground avoidance. Nevertheless, the relatively gradual degradation of the S/N as $\sigma_z$ and $m$ increase from their fiducial values suggests that the signal is still detectable for less optimistic assumptions about photo-$z$ errors and foreground avoidance. This reinforces the viability of our higher-order statistic as a practical means to cross-correlate EoR galaxies lacking spectroscopic redshift precision with 21\,cm observations. 

Consistent with the results shown in figure~\ref{fig:errbudget}, the right column of figure~\ref{fig:survey} confirms that increasing the overlapping survey area generally provides substantially greater leverage than increasing the survey integration time per pointing of the 21\,cm survey. At fixed depth, the S/N grows approximately as $A_{\rm tot}^{1/2}$, as expected from the increase in the number count of independent Fourier modes. Increasing $t_{\rm field}$ for SKA-Low observations, by contrast, produces a much weaker improvement and begins to saturate after the thermal contribution becomes subdominant. Under these assumptions, expanding the common footprint for Roman and SKA-Low, which is permitted by Roman's 19.2\,deg$^2$ Deep-Tier HLWAS observations, is therefore a more effective way to improve the detectability than dedicating substantially more SKA-Low integration time to the same survey area. At smaller scales ($k \gtrsim 0.2\,h\mathrm{Mpc}^{-1}$) and higher redshifts, deeper Roman observations may help improve the measurements by increasing the number density of detectable LBGs and thereby reducing the shot noise (figure~\ref{fig:errbudget}). Such opportunities might be pursued through coadded imaging from Roman's High-Latitude Time-Domain Survey (HLTDS; \citep{RomanSurveys2025}) or by dedicated future General Investigator programs, given appropriate filter coverage, depth, and overlap with SKA-Low observations.

\subsubsection{Constraints on physical parameters}

We next use the cross-power spectrum to forecast constraints on the four astrophysical parameters
$\boldsymbol{\theta} = \left\{
f_{\rm esc},
\log(L_X/\mathrm{SFR}),
\alpha_\eta,
\log(T_{\rm vir,min}/\mathrm{K})
\right\}$
around the fiducial values $f_{\rm esc} = 0.1$, $\log(L_X/\mathrm{SFR}) = 40.5$, $\alpha_\eta=0.67$, $\log(T_{\rm vir,min}/\mathrm{K})=4.0$ motivated by observational constraints on the reionization history \citep{Jones2025,Kageura2025,Mason2026,Umeda2026} and the star formation efficiency of EoR galaxies \citep{SF2016,Mirocha2017,SippleLidz2024}. We combine the measurements at $z=6.34$ and 7.77, which represent two well separated, significantly detected cross-correlations with opposite signs and substantially different mean neutral fractions ($\langle x_{\rm HI} \rangle = 0.13$ and 0.72, respectively). The Fisher derivatives are evaluated using the converged central-difference offsets, with the filtering, covariance, and accessible Fourier modes being held fixed across each parameter variation.

\begin{figure*}[!ht]
 \centering
 \includegraphics[width=\textwidth]{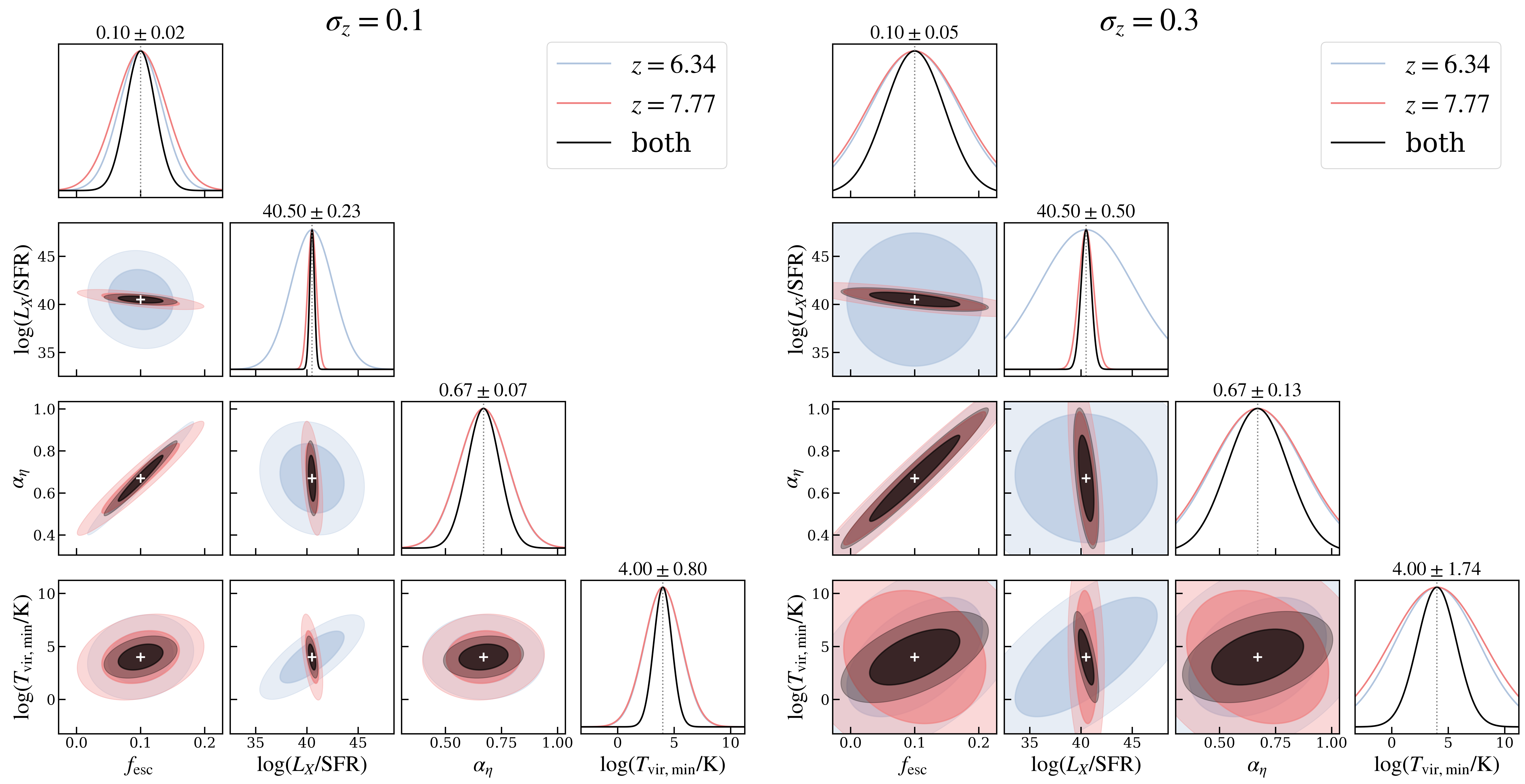}
 \caption{\textit{Left:} Joint 68\% and 95\% CL constraints on the escape fraction of ionizing photons ($f_\mathrm{esc}$), the X-ray heating efficiency ($L_X$), the stellar feedback efficiency ($\alpha_{\eta}$), and the minimum virial temperature for star-forming halos ($T_\mathrm{vir,min}$) obtained from the Fisher matrix analysis of LBG--21\,cm$^2$ cross-power spectra detectable by Roman and SKA-Low at $z = 6$--8. The constraints from $z=6.34$ where the cross-correlation is negative and $z=7.77$ where it is positive are complementary, whose combination reduces parameter degeneracies especially for $L_X$. \textit{Right:} The same as the left panel but for the case of a larger photo-$z$ error $\sigma_z=0.3$.}
 \label{fig:fisher}
\end{figure*}

As shown by the resulting parameter ellipses in figure~\ref{fig:fisher}, the two redshifts provide complementary information. Despite its lower predicted S/N, the late-EoR signal at $z=6.34$ gives slightly tighter constraints on $f_{\rm esc}$, consistent with the strong response of the negative cross-power spectrum to changes in ionizing efficiency at this stage. However, this redshift alone provides only weak constraints on the X-ray luminosity responsible for IGM heating. The earlier $z=7.77$ signal retains greater sensitivity to the thermal history, thereby producing much tighter constraints on $L_X$. For the marginalized joint distribution of $\alpha_\eta$ and $T_\mathrm{vir,min}$, the comparison between left and right panels also reveals that $\sigma_z$ can modify the relative weighting of available modes and thereby change the complementarity between measurements at different redshifts. 

Combining constraints from the two redshifts yields absolute 1-$\sigma$ uncertainties of 
$\sigma(\boldsymbol{\theta}) = \{0.02,\ 0.23,\ 0.07,\ 0.80\}$ for $\sigma_z = 0.1$ and $\sigma(\boldsymbol{\theta}) = \{0.05,\ 0.50,\ 0.13,\ 1.74\}$ for $\sigma_z = 0.3$. The improvement arises not only from the increased signal-to-noise ratio, but also from the different orientations of the parameter degeneracies at the two redshifts. In particular, $f_{\rm esc}$ and $\alpha_\eta$ exhibit a strong positive covariance at either redshift. Increasing $\alpha_\eta$ suppresses star formation more strongly in low-mass halos and tends to delay reionization, while increasing $f_{\rm esc}$ raises the ionizing output per unit star formation and compensates for this effect. Models with larger $\alpha_\eta$ therefore generally require larger $f_{\rm esc}$ to preserve a similar ionization history and cross-correlation signal.

These constraints should be interpreted as forecasts based on linearization around the fiducial parameter values within the adopted \texttt{LIMFAST} model and observational assumptions. Nevertheless, they demonstrate that the redshift evolution of the LBG--21\,cm$^2$ cross-spectrum contains information beyond its overall detectability. In particular, combining measurements from distinct stages of reionization can separate physical parameters that would remain highly degenerate at either redshift alone. The same modeling caveats apply to other 21\,cm experiments, though the attainable sensitivity and accessible Fourier modes depend on the instrumental configuration. To provide a complementary example in addition to SKA-Low, we repeat the detectability and parameter constraint analysis for the Hydrogen Epoch of Reionization Array (HERA) and discuss the results in appendix~\ref{sec:hera}.

\subsection{Interpreting the redshift evolution of the LBG--21\,cm$^2$ cross-correlation} \label{sec:results:interpret}

The sign evolution of the LBG--21\,cm$^2$ cross-correlation revealed by figure~\ref{fig:constraints} indicates a qualitative change in how the 21\,cm fluctuation power surviving the filtering is distributed around galaxies. To examine this behavior, we use two complementary diagnostics. First, we apply a three-dimensional Morlet wavelet decomposition to examine whether 21\,cm fluctuation power on different filtered scales is locally enhanced or suppressed around galaxy overdensities. Second, we decompose the 21\,cm brightness temperature field into neutral fraction, density, and other components and propagate their pairwise products through the same filtered-then-squared cross-correlation analysis.

For the wavelet analysis, we define a localized 21\,cm power field using the convolution with the Morlet wavelet filter
\begin{equation}
\mathcal{E}_q(\mathbf{x})
=
\left\langle
\left|
\psi_{q,\hat{\mathbf n}}
*
\mathcal{T}
\right|^2
\right\rangle_{\hat{\mathbf n}},
\label{eq:morlet_energy}
\end{equation}  
where $\mathcal{T}$ is the fully filtered 21\,cm field, $\tilde{\psi}_{q,\hat{\mathbf n}} \propto \exp \left[ -|\boldsymbol{k}-q \hat{\boldsymbol{n}}|^2 / (2 f^2 q^2) \right]$ is a Morlet packet centered on the wavevector $q\hat{\mathbf n}$, and the average is taken over the survey specific orientations. We inspect $q=0.2$, 0.4, 0.6, 0.8, and $1.0\,h\,{\rm Mpc}^{-1}$ with a width of $f=0.3$. The Morlet packet has the advantage of being localized in both configuration and Fourier space, which allows $\mathcal{E}_q(\mathbf{x})$ to map the local fluctuation power of the filtered 21\,cm field in a range of wavenumbers around $q$ throughout the simulation volume. The width parameter $f$ controls the localization trade-off by specifying a characteristic width $fq$ of the Fourier-space Gaussian packet and another characteristic width $\sim(fq)^{-1}$ of the configuration-space envelope. Increasing $f$ therefore improves spatial localization, whereas decreasing $f$ selects a narrower range of wavenumbers around $q$ while averaging over a larger spatial region. 

Computing this field by convolution over the full periodic volume avoids introducing the sharp internal boundaries associated with division into separate, nonperiodic subvolumes (e.g., \citep{Sun2025}), which may lead to artifacts in local power estimates. The cross-correlation coefficient, $r_{\mathrm{gal}\times\mathcal{E}_q}(k)$, then reveals how the 21\,cm power on each selected scale $q$ correlates with galaxy overdensities. Here, $q$ specifies the 21\,cm fluctuations whose power is measured, whereas $k$ describes the scale of their spatial modulation. Comparing the correlations across $q$ therefore identifies which fluctuation scales exhibit enhanced or suppressed power in galaxy overdensities. This helps interpret the sign evolution of the LBG--21\,cm$^2$ cross-correlation. Although used here mainly to build intuition for our higher-order statistic, the Morlet wavelet transform has been previously shown to be a powerful approach for extracting localized reionization information from tomographic 21\,cm data \citep{Trott2016}. 

\begin{figure*}[!ht]
 \centering
 \includegraphics[width=\textwidth]{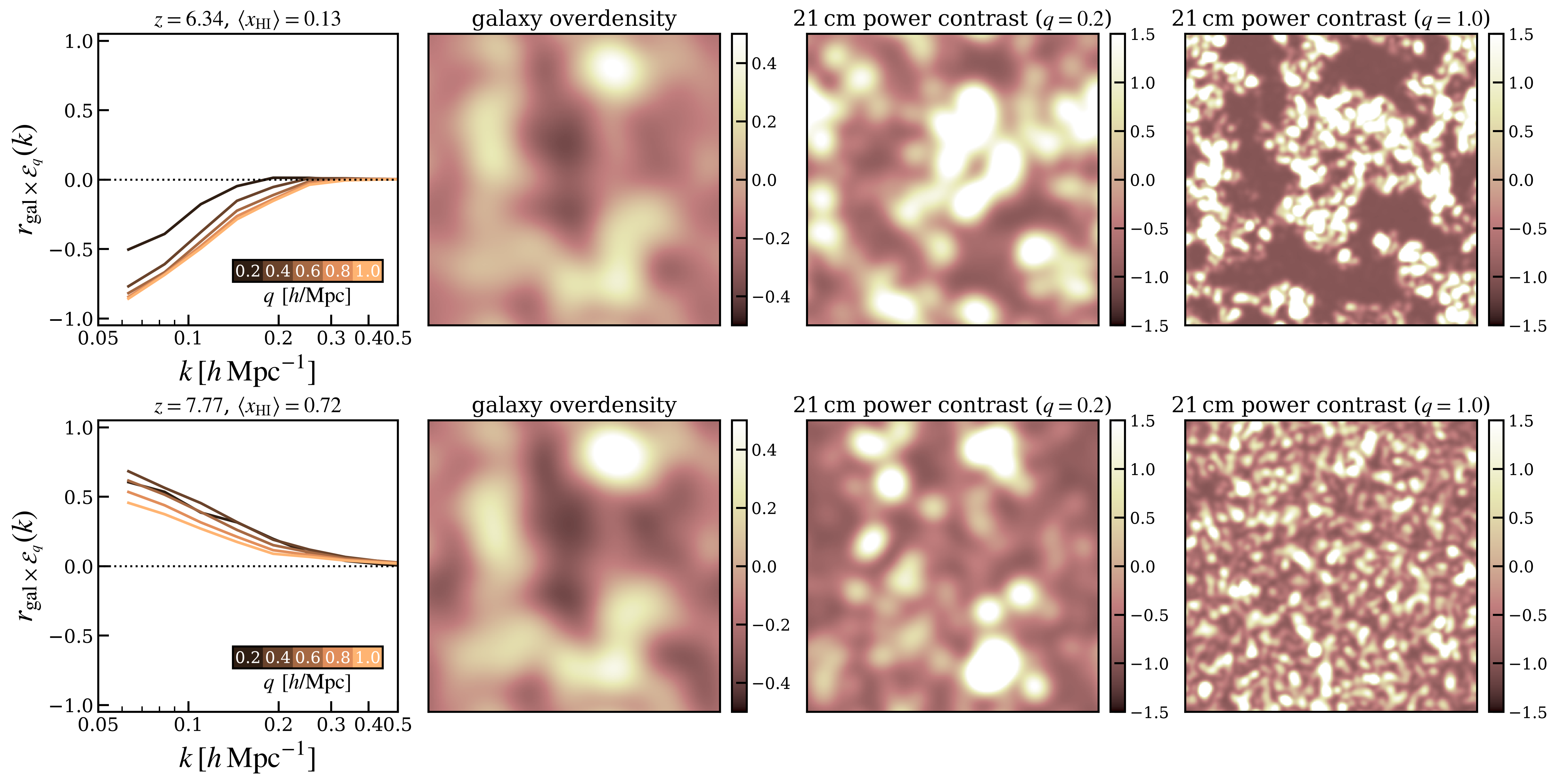}
 \caption{Morlet wavelet decomposition of the foreground-filtered 21\,cm fluctuations around overdensity environments traced by Roman LBGs. Wavelet filters centered at $q=0.2$--$1.0\,h\,{\rm Mpc}^{-1}$ are applied to construct the localized 21\,cm power field $\mathcal{E}_q$. The left panels show the cross-correlation coefficients, $r_{{\rm gal}\times\mathcal{E}_q}$, at $z=6.34$ (top) and $z=7.77$ (bottom) where the LBG--21\,cm$^2$ cross-correlation signals are strongly negative and positive, respectively. The remaining panels show central slices of the smoothed galaxy overdensity and the 21\,cm power contrasts, $\mathcal{E}_q/\langle\mathcal{E}_q\rangle-1$, localized at $q=0.2$ and $1.0\,h\,{\rm Mpc}^{-1}$. The negative correlations at $z=6.34$ indicate that the surviving 21\,cm fluctuations are locally suppressed around Roman LBGs at the late EoR, whereas the positive correlations at $z=7.77$ suggest locally enhanced 21\,cm power around Roman LBGs at an earlier stage of the EoR.}
 \label{fig:morlet}
\end{figure*}

Figure~\ref{fig:morlet} shows that 21\,cm power fields localized by the Morlet wavelet filter follow qualitatively different environments at the two redshifts examined. At $z=6.34$, $r_{{\rm gal}\times\mathcal{E}_q}$ is negative on large scales for all five wavelet scales. The anti-correlation becomes stronger for the higher-$q$ filters, indicating that small-scale 21\,cm fluctuation power is particularly suppressed around the galaxy population. This is consistent with the late-EoR ($\langle x_{\rm HI}\rangle\simeq0.13$) morphology, in which LBGs reside inside extended ionized regions and trace large-scale overdensities, with brightness temperature fluctuations of the surrounding neutral IGM removed. At $z=7.77$, the wavelet correlations are instead positive. The strongest positive correlation occurs for the lowest-$q$ filter, whereas the correlation strength decreases progressively toward higher $q$. At this earlier stage ($\langle x_{\rm HI}\rangle\simeq0.72$), overdense regions traced by LBGs retain substantial intermingled patches of ionized and neutral gas. These regions consequently exhibit enhanced localized 21\,cm fluctuation power rather than the extended suppression characteristic of the later stages of reionization. The reduced positive correlation as $q$ increases is consistent with ionized patches already suppressing finer-scale 21\,cm fluctuations around LBGs. Similar behavior is revealed by position-dependent 21\,cm power spectra, where inside-out reionization suppresses 21\,cm power on growing scales in overdense environments as ionized regions expand and merge \citep{Giri2019}. In both cases, $r_{{\rm gal}\times\mathcal{E}_q}$ approaches zero at high $k$, showing that the galaxy modulation of the local 21\,cm power is primarily a large-scale environmental effect.

\begin{figure*}[!ht]
 \centering
 \includegraphics[width=\textwidth]{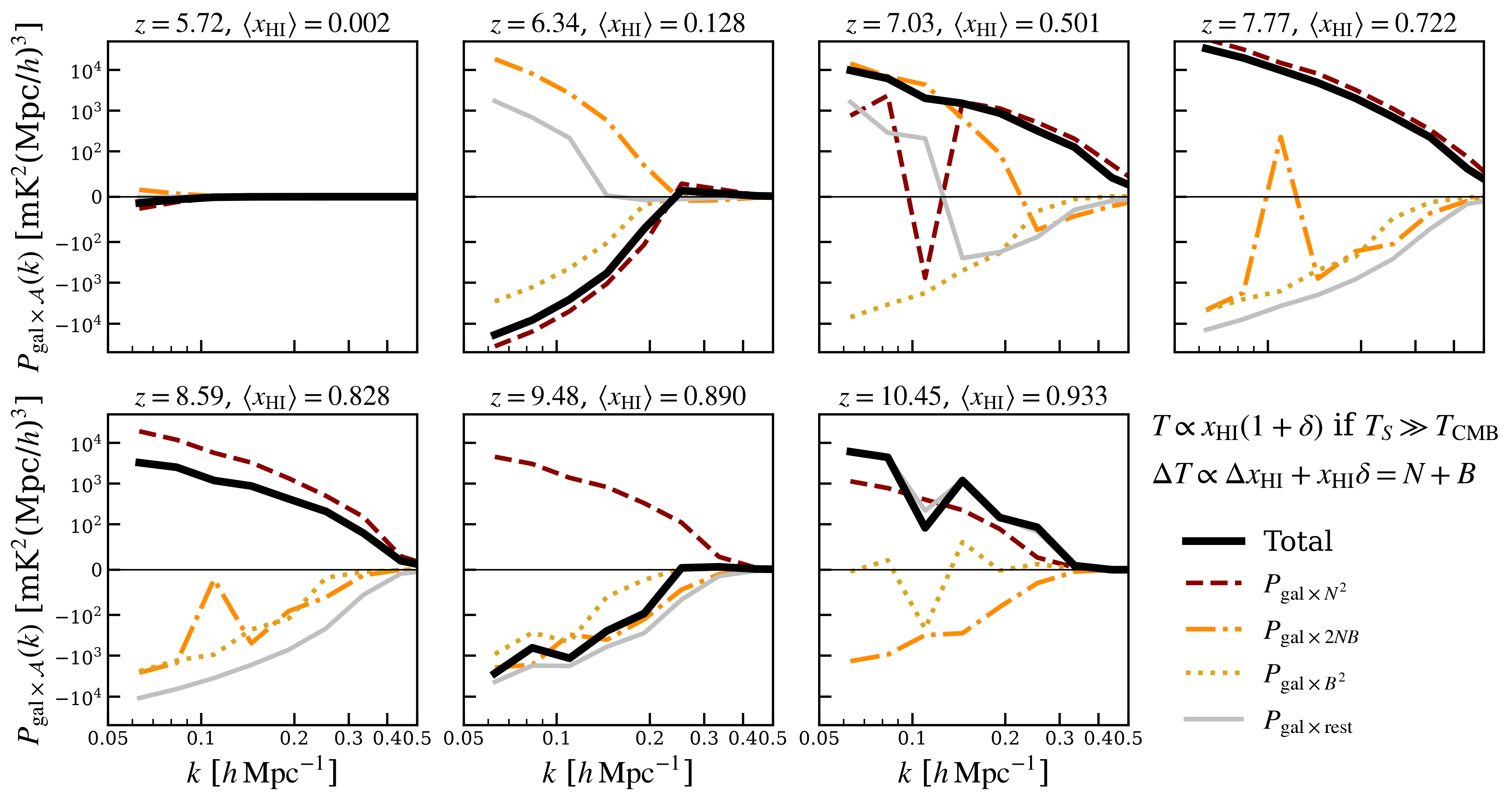}
 \caption{Decomposition of physical contributions to the LBG--21\,cm$^2$ cross-power spectrum across the EoR. Each panel shows the total signal (black solid) together with contributions from the neutral fraction ($P_{\mathrm{gal} \times N^2}$; red dashed), the neutral-fraction-weighted density ($P_{\mathrm{gal} \times B^2}$; yellow dotted), the neutral fraction--density cross term ($P_{\mathrm{gal} \times 2NB}$; orange dashed-dotted), and the remainder that collects all contributions not accounted for ($P_{\mathrm{gal} \times \mathrm{rest}}$; gray solid), including correlations involving spin temperature fluctuations and the residual of the brightness temperature reconstruction. The $P_{\mathrm{gal} \times N^2}$ contribution drives the robust transition from negative correlation at $z\lesssim6$ to positive correlation at $z \sim 7$--8. At still higher redshifts ($z \gtrsim 9$) where the total signal becomes weak, the remaining thermal and residual contributions become significant compared to $P_{\mathrm{gal} \times N^2}$, producing more complex behavior. The neutral fraction evolution therefore serves as the primary, though not exclusive, driver for the sign evolution of the LBG--21\,cm$^2$ cross-correlation.}
 \label{fig:components}
\end{figure*}

In figure~\ref{fig:components}, we inspect how the evolution of the LBG--21\,cm$^2$ signal is driven by the contributions from different components of the 21\,cm fluctuations. When the EoR occurs in the saturated spin temperature limit ($T_S \gg T_\mathrm{CMB}$), the 21\,cm brightness temperature fluctuations can be decomposed into fluctuations in the neutral fraction itself and density fluctuations weighted by it, namely $\Delta T \propto \Delta x_\mathrm{HI} + \delta \langle {x}_\mathrm{HI} \rangle + \delta \Delta x_\mathrm{HI} = \Delta x_\mathrm{HI} + \delta x_\mathrm{HI} = N + B$. For this decomposition, we apply the same foreground and Wiener filters $W_{21}$ to each component of the 21\,cm field before squaring or multiplying by other components. Each resulting product is then cross-correlated with the galaxy field with photo-$z$ filter $W_{\rm gal}$ included. 

At $z=6.34$, the $P_{\mathrm{gal} \times N^2}$ contribution is strongly negative and dominates the total cross power due to the suppression of neutral fraction fluctuations around bright Roman-detectable galaxies. The positive $P_{\mathrm{gal} \times 2NB}$ contribution partially counteracts this signal, while $P_{\mathrm{gal} \times B^2}$ and the remaining terms provide smaller corrections. The transition at $z=7.03$ is scale-dependent in that the $P_{\mathrm{gal} \times N^2}$ contribution does not universally set the sign of the total cross power. While it can drop to negative on large scales, the $P_{\mathrm{gal} \times 2NB}$ terms remain strongly positive, producing the more subtle yet still positive total cross power. This is reasonable for an intermediate stage with $\langle x_{\rm HI}\rangle \simeq 0.5$, when the interplay between neutral fraction and density fields can contribute significantly. At $z=7.77$ and 8.59, the total large-scale cross-power is positive, with the $P_{\mathrm{gal} \times N^2}$ contribution providing the largest positive contribution. The density and remaining thermal terms partially counteract it but do not reverse the total sign. These redshifts therefore represent the clear regime where enhanced neutral fraction fluctuations around galaxy overdensities produce a positive LBG--21\,cm$^2$ cross-correlation, consistent with the wavelet analysis shown in figure~\ref{fig:morlet}. At $z\gtrsim9$, the total signal becomes weak and the interpretation is less accurately described by the $P_{\mathrm{gal} \times N^2}$ term alone. In particular, at $z=9.48$, the $P_{\mathrm{gal} \times N^2}$ term remains positive, but negative contributions from other terms cause the total signal to become slightly negative. At $z=10.45$, the remaining terms become significant enough and yield a weakly positive total signal. These cases explain why the highest-redshift correlations in figure~\ref{fig:constraints} are weak and why their signs do not necessarily follow $P_{\mathrm{gal} \times N^2}$. The non-monotonic evolution over $10.5 \gtrsim z \gtrsim 8.5$ ($0.07 \lesssim x_{\rm HII} \lesssim 0.17$) further hints at the interplay of multiple drivers of the 21\,cm signal near the onset of reionization when ionized bubbles first start to form (see, e.g., \citep{Libanore2026} for a recently proposed probe of this era) . We also note that, unlike other components, $P_{\mathrm{gal} \times B^2}$ does not dominate the total cross power but remains predominantly negative throughout, indicating persistently suppressed power of $x_\mathrm{HI}$-weighted density fluctuations around galaxy overdensities. 


Together, figures~\ref{fig:morlet} and \ref{fig:components} suggest that the sign evolution of the LBG--21\,cm$^2$ cross-correlation is primarily, though not exclusively, driven by the evolving IGM neutral fraction. The growth of ionized regions changes the local 21\,cm fluctuation power around galaxies from enhanced at the earlier EoR to suppressed at the later EoR, producing the characteristic positive-to-negative transition. Density modulations and spin temperature fluctuations can introduce physically meaningful corrections and become especially important near the transition and at the highest redshifts.

\section{Discussion and Conclusions} \label{sec:conclusions}

In this work, we show that the apparent loss of synergy between foreground-filtered 21\,cm data and photometrically selected LBGs during the EoR is not unavoidable. In particular, filtering and then squaring the 21\,cm field before correlating with LBGs restores a non-vanishing and physically informative signal through mode coupling induced by higher-order correlations. Using \texttt{LIMFAST} simulations over $6 \lesssim z \lesssim 10$, we find that this LBG--21\,cm$^2$ cross-correlation signal can be detected at high significance across much of the EoR by combining a deep Roman LBG sample with $m_\mathrm{AB,lim}=27$ and modestly deep ($\sim$500 hours per field) SKA-Low observations. Although the exact constraints on the EoR physics are model dependent, our Fisher matrix analysis suggests that measurements of this statistic can provide valuable insights into the ionization and thermal histories of the IGM and the underlying ionizing source populations that complement auto-correlation measurements. It is therefore promising to leverage this higher-point cross-correlation to unlock the synergy between forthcoming large 21\,cm and photometric galaxy surveys of the EoR. 

Our statistic is closely connected to, but physically distinct from, the usual galaxy--21\,cm cross-power spectrum that has been studied extensively in the literature. In the saturated spin temperature ($T_S \gg T_\mathrm{CMB}$) limit, the direct cross-power is expected to evolve from positive at early times, when galaxies trace overdense and 21\,cm-bright regions, to negative once their surroundings become preferentially ionized \citep{Lidz2009}. Before heating saturates, however, $T_S$ fluctuations can substantially complicate this evolution. In particular, \cite{Moriwaki2024} showed that the positive-to-negative transition in their models occurs only after the mean $T_S$ exceeds the CMB temperature \emph{and} ionization fluctuations become more important than $T_S$ fluctuations. The sign of the direct cross-power spectrum alone therefore does not, in general, provide an unambiguous determination of whether the global 21\,cm signal is in emission or absorption \citep{Heneka2020}. The sign of our statistic has a different interpretation: a positive (negative) LBG--21\,cm$^2$ cross-power signal indicates that the local small-scale 21\,cm variance is enhanced (suppressed) around galaxy overdensities. Given that our squared statistic is invariant under $\delta T_{\rm b} \rightarrow -\delta T_{\rm b}$, it likewise cannot distinguish global emission from absorption. In addition to its main observational advantage of probing the coupling between galaxies and local 21\,cm power even when the smoothing by photo-$z$ uncertainties and the filtering of 21\,cm foregrounds leave little overlapping Fourier space, it naturally complements the information from direct cross-correlations using spectroscopic galaxy samples \citep{LaPlante2023,Moriwaki2024,Gagnon-Hartman2025}. Being sensitive to environmental modulations of 21\,cm fluctuations when the conditional mean is weak or cancels, the LBG--21\,cm$^2$ correlation can be viewed as a projection of the underlying galaxy--21\,cm--21\,cm cross-bispectrum, closely related to position-dependent power spectrum statistic \citep{Chiang2014,Giri2019,Sun2025}. Combining many triangle configurations into the lower dimension cross-power spectrum simplifies the analysis and offers better interpretability, but this does not retain the full information from the cross-bispectrum.

A number of caveats and possible extensions are noteworthy for the proof-of-concept analysis presented here. First, observational effects on the simulated signals are modeled in an intentionally simplified way. In practice, details of the accessible Fourier modes and their covariance will depend on many additional factors, including the survey geometry, instrumental response, detailed foreground mitigation and power spectral analysis strategies, and fidelity of the photo-$z$ reconstruction \citep{LiuShaw2020,NewmanGruen2022ARA&A}. A more realistic treatment of these transfer functions will therefore be essential for delivering more realistic survey-specific forecasts. Second, the present forecasts are based on a particular astrophysical model that is broadly consistent with currently available EoR constraints. While physically grounded, this fiducial case represents only one of many possible reionization scenarios allowed by existing observations. In addition, modeling uncertainties extend beyond the values of parameters within a single model. For instance, choices of the halo mass function and stellar population synthesis model can alter the inferred ionization and thermal histories \citep{Mirocha2021a}, whereas variations in galaxy assembly histories can alter the connection between observed galaxies and the ionized regions around them \citep{Mirocha2021b}. Recent studies have also suggested that the implementation of the ionizing photon mean free path can substantially affect the reionization history, morphology, and their imprints on the 21\,cm signal, particularly toward the end of reionization \citep{Davies2022,Wyatt2026}. These effects are especially relevant to the LBG--21\,cm$^2$ statistic studied, which depends not only on the global ionization and thermal histories but also on how the 21\,cm topology connects to the observed galaxies. Finally, the amount of three-point information preserved by this projection has not yet been quantified. Several previous studies of the 21\,cm bispectrum have demonstrated that its scale, sign, and shape dependence reveal valuable information about Ly$\alpha$ coupling, X-ray heating, and ionization topology and analyzing different triangle configurations helps tighten the constraints on EoR parameters \citep{Shimabukuro2016,Hutter2020,Tiwari2022}. Because our squared field integrates over pairs of filtered 21\,cm modes, physically distinct contributions can mix together and partially cancel. The parameter constraints presented are therefore only part of the full three-point information retained by our statistic.

These limitations point toward several promising directions for future work. On the modeling side, it will be useful to replace the simple foreground and photo-$z$ window functions with realistic, redshift-dependent transfer functions tailored to more specific instrument and survey configurations. Meanwhile, a larger suite of simulations spanning alternative ionizing source and sink prescriptions, along with a broader parameter space, will ultimately be needed to refine our quantitative predictions and their physical interpretation. On the higher-order statistics, a useful next step would be a detailed comparison between the LBG--21\,cm$^2$ statistic and the full galaxy--21\,cm--21\,cm cross-bispectrum. Such an analysis can quantify the trade-off between the practical gain from the compact statistic and the information lost by integrating over triangle configurations. Although previous studies have considered the EoR 21\,cm--[CII]--[CII] \citep{Beane2018} and post-reionization 21\,cm--galaxy \citep{Noble2026} cross-bispectra, a full cross-bispectrum analysis tailored to LBGs and foreground-filtered 21\,cm signal during the EoR has not yet been carried out. More broadly, our results suggest that higher-order statistics may offer a practical means of recovering otherwise inaccessible cross-correlation information when one tracer loses low-$k_\parallel$ modes while the other loses high-$k_\parallel$ modes. Indeed, several recent studies have explored related applications involving three-dimensional 21\,cm maps and angular fluctuations of the CMB and CIB \citep{Sun2025,Zhou2025,Yuwen2026}. From a theoretical perspective, extending the perturbative bias expansion of the 21\,cm field \citep{McQuinnDAloisio2018,Qin2022,Kokron2025} to our higher-order statistic would help clarify the physical origins of the sign evolution and scale dependence of the LBG--21\,cm$^2$ signal, thereby strengthening its potential as a powerful probe of the EoR. Since the removal of spectrally smooth continuum foregrounds is also a necessary component of analysis for other LIM signals (though generally less challenging for target lines such as [CII] and CO than for 21\,cm), the statistic considered here may have useful applications beyond 21\,cm surveys. One example is the cross-correlation of high-$z$ [CII] LIM signals with narrowband-selected LAEs, which serves as a useful validation of [CII] auto-correlation analysis and helps constrain physical properties of [CII] emitters \citep{Sun2021}. 

In summary, we find that 21\,cm and photometric galaxy surveys can still be effectively combined through the LBG--21\,cm$^2$ cross-correlation. This statistic preserves the convenience of a two-point power spectrum while capturing higher-order information that survives the observational mode mismatch strongly suppressing the direct cross-power spectrum. Our results therefore highlight a promising avenue for jointly exploiting forthcoming Roman and SKA-Low observations of the EoR and illustrate the broader potential of higher-order cross-correlations as probes of the early Universe.

\appendix

\section{Detectability of the Cross-correlation Coefficient} \label{sec:ccc}

In the main text, we base our predicted EoR constraints on the cross-power spectrum itself, instead of the cross-correlation coefficient whose inference more heavily depends on the two auto-correlations involved. For completeness, we describe here how the detectability of the cross-correlation coefficient can be estimated using the Fisher matrix formalism. For the cross-correlation coefficient
\begin{equation}
r_{\mathrm{gal}\times\mathcal{A}}(k, \mu) = \frac{P_{\mathrm{gal}\times\mathcal{A}}(k, \mu)}{\sqrt{P_\mathrm{gal}(k, \mu) P_\mathcal{A}(k, \mu)}},
\end{equation}
we can define the covariance matrix as
\begin{equation}
C(k, \mu) = \left[ \begin{matrix}
\delta P_\mathcal{A}(k, \mu) & r(k, \mu) \sqrt{P_\mathcal{A}(k, \mu)P_\mathrm{gal}(k, \mu)} \\
r(k, \mu) \sqrt{P_\mathcal{A}(k, \mu)P_\mathrm{gal}(k, \mu)} & \delta P_\mathrm{gal}(k, \mu)
\end{matrix} \right]
\end{equation}
For the observables $r$, $P_\mathcal{A}$, and $P_\mathrm{gal}$, we can express the full Fisher matrix in a given $k$ bin as
\begin{equation}
F_{ij}(k) = \int_0^1 d \mu \frac{k^2 \Delta k V_\mathrm{survey}}{4\pi^2} f_{ij}(k, \mu), 
\end{equation}
where $f_{ij}$ is a $3 \times 3$ matrix given by
\begin{equation}
f_{ij} = \frac{1}{2} \mathrm{Tr}(C^{-1} C_{,i} C^{-1} C_{,j}), \hspace{0.3cm} \mathrm{with}\ i,j \in \{r, P_\mathcal{A}, P_\mathrm{gal} \}
\end{equation}
The resulting variance of $r$ after marginalizing over $P_\mathcal{A}$ and $P_\mathrm{gal}$ is then
\begin{equation}
[\delta r(k)]^2 = \left[ F^{-1}(k) \right]_{rr}. 
\end{equation}

\section{UVLFs and Clustering of EoR Galaxies Predicted by \texttt{LIMFAST}} \label{sec:uvlf}

\begin{figure*}
 \centering
 \includegraphics[width=\textwidth]{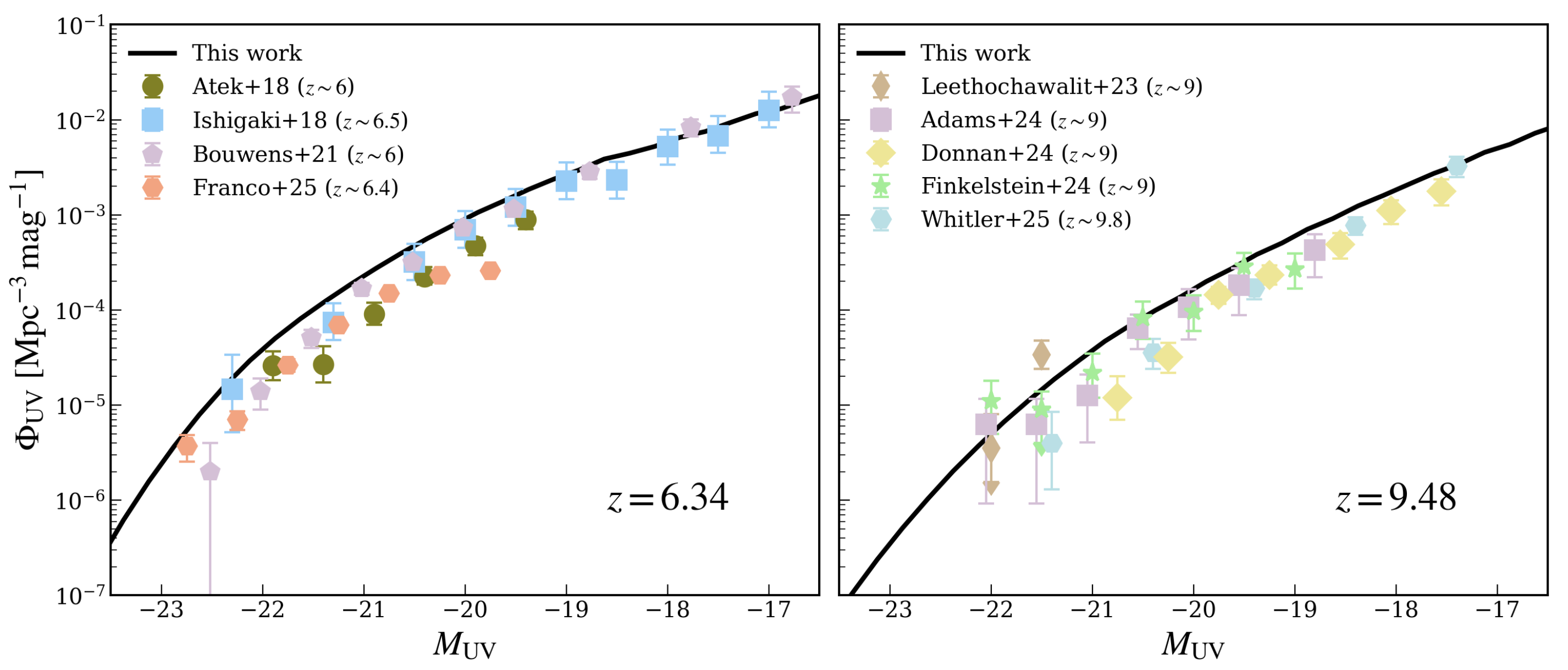}
 \caption{Comparison of the $z\sim6$--7 and $z\sim9$--10 UVLFs predicted by the galaxy formation model in \texttt{LIMFAST} (after dust attenuation \citep{ZhaoFurlanetto2024}) against observations from HST \citep{Atek2018,Ishigaki2018,Bouwens2021,Leethochawalit2023} and JWST \citep{Adams2024,Donnan2024,Finkelstein2024,Franco2025,Whitler2025}. The scatter among published measurements highlights the non-negligible observational uncertainties, including contributions from cosmic variance.}
 \label{fig:uvlf}
\end{figure*}

The \texttt{LIMFAST} simulations used to predict the LBG and 21\,cm signals throughout this work assume the galaxy formation model with energy-conserving stellar feedback \citep{Furlanetto2017}, which is calibrated against high-$z$ UVLFs observed by HST and JWST. While recent JWST observations suggest that SFR stochasticity plays a significant role in shaping the bright-end UVLF at high redshifts \citep{Shen2023,Sun2023ApJL,Sun2025JCAP,Munoz2026}, this effect is most pronounced at $z>10$. Since our EoR analysis focuses primarily on $6 \lesssim z \lesssim 10$, we defer the exploration of SFR stochasticity to future work. Figure~\ref{fig:uvlf} demonstrates the broad agreement between the UVLF predictions from \texttt{LIMFAST} and observations at $z\sim6$--7 and $z\sim9$--10. Dust attenuation is applied following the prescription in \citep{ZhaoFurlanetto2024}, which mainly affects the bright end and diminishes at $z\gtrsim9$. Specifically, we model the mean UV slope as
\begin{equation}
\beta = \frac{d\beta}{dM^{\rm obs}_{\rm UV}} \left( M_{\rm UV}^{\rm obs} - M_0 \right) +\beta_{M_0},
\label{eq:beta_Muv}
\end{equation}
where $\beta_{M_0}=-0.081z-1.58$, $d\beta/dM^{\rm obs}_{\rm UV}=0.012z-0.216$, and $M_0=-19.5$. Combining eq. (\ref{eq:beta_Muv}) with the Meurer relation for dust correction $A_{\rm UV}=4.43 + 1.99\beta$ \citep{Meurer1999} and the definition $M_{\rm UV}^{\rm obs}=M_{\rm UV}^{\rm int}+A_{\rm UV}$ gives
\begin{equation}
M_{\rm UV}^{\rm obs}
=
\frac{
M_{\rm UV}^{\rm int}
+4.43+1.99 \beta_{M_0}
-1.99 (d\beta/dM^{\rm obs}_{\rm UV}) M_0
}{
1 - 1.99 d\beta/dM^{\rm obs}_{\rm UV}
}.
\end{equation}
We require non-negative attenuation and further include a Gaussian scatter of $\sigma_{A_{\rm UV}}=0.30$ mag around the mean $M_{\rm UV}^{\rm obs}$.

\begin{figure*}
 \centering
 \includegraphics[width=0.8\textwidth]{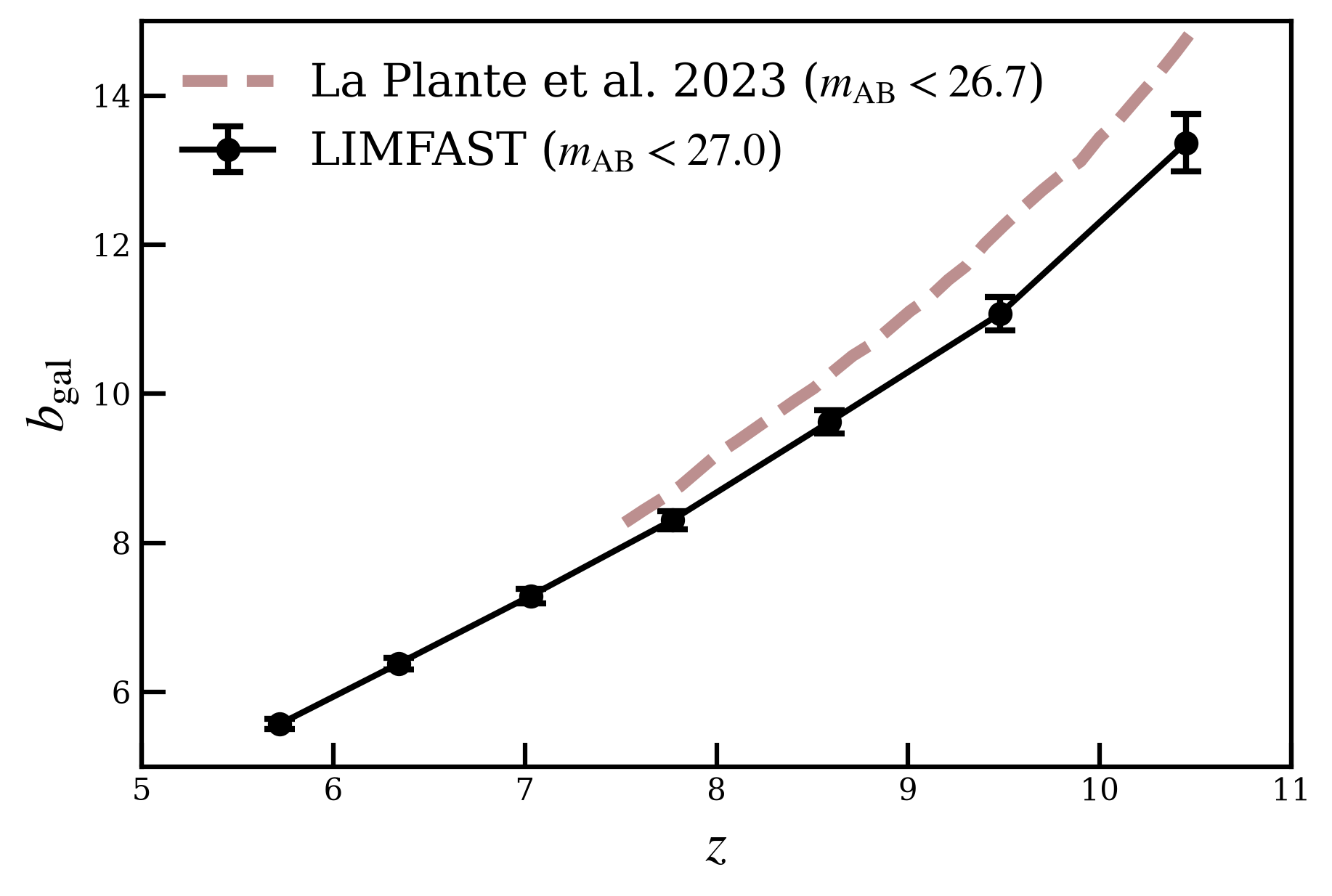}
 \caption{Clustering bias and its 1$\sigma$ fitting uncertainty of galaxies with $m_\mathrm{AB}<27$ (comparable to the limiting magnitude of Roman HLWAS Deep-Tier observations) as a function of redshift, measured by fitting a linear bias model $P_\mathrm{gal \times m} = b_\mathrm{gal} P_\mathrm{mm}$ to the distributions of LBGs and matter overdensities simulated by \texttt{LIMFAST}. For comparison, we also show bias estimates for $m_\mathrm{AB}<26.7$ galaxies from \citep{LaPlante2023}, based on the linear halo bias \citep{Tinker2010} and semi-empirically predicted galaxy properties \citep{Mirocha2020}.}
 \label{fig:bias}
\end{figure*}

In addition to reproducing the observed UVLFs, reliable predictions of the cross-correlation require realistic spatial clustering of the simulated galaxies. We quantify the large-scale clustering of our simulated Roman LBG sample by fitting the linear bias model
\begin{equation}
P_{\rm gal\times m}(k)=b_{\rm gal}P_{\rm mm}(k). 
\end{equation}
We choose the cross-power spectrum because it contains no additive galaxy shot noise. This makes it a more direct bias estimator than the galaxy auto-power spectrum. As illustrated in figure~\ref{fig:bias}, the predicted $b_\mathrm{gal}$ increases from about 5.5 at $z=5.72$ to 14 at $z=10.45$, owing to the increasingly rare host halos selected by the condition $m_{\rm AB} < 27$ at higher redshifts. The evolution agrees broadly with the prediction of \citep{LaPlante2023}, with the slightly lower normalization qualitatively expected from the deeper limiting magnitude in our case. In terms of the validity of a scale-independent $b_\mathrm{gal}$, we verify that the inferred bias factor remains constant to $k \simeq 0.3\,h\,\mathrm{Mpc}^{-1}$. The rms fractional variations around the best-fit constant remain below 5\% through $z\sim8$ and grows to about 10\% at $z\sim10$, by which the increasingly sparse galaxy samples start to introduce substantial shot noise. Therefore, a linear bias is reasonable on large scales over the redshift range considered. Nevertheless, using resolved halo fields from mini-Uchuu preserves the information about rare peaks and non-linearity that would be absent from a purely linear bias prescription. We also use the estimated $b_\mathrm{gal}$ to characterize the scale at which galaxy shot noise starts to dominate. Given $m_{\rm AB} < 27$, we find that at $z \sim 6$, 7, and 9.5 the shot noise power exceeds the clustering power at $k \gtrsim 0.5$, 0.3, and $0.1 h\,\mathrm{Mpc}^{-1}$, respectively. This shift toward lower $k$ scales as redshift increases reflects the decreasing number density of Roman-detectable LBGs visualized in figure~\ref{fig:signals}. 

\begin{figure*}
 \centering
 \includegraphics[width=\textwidth]{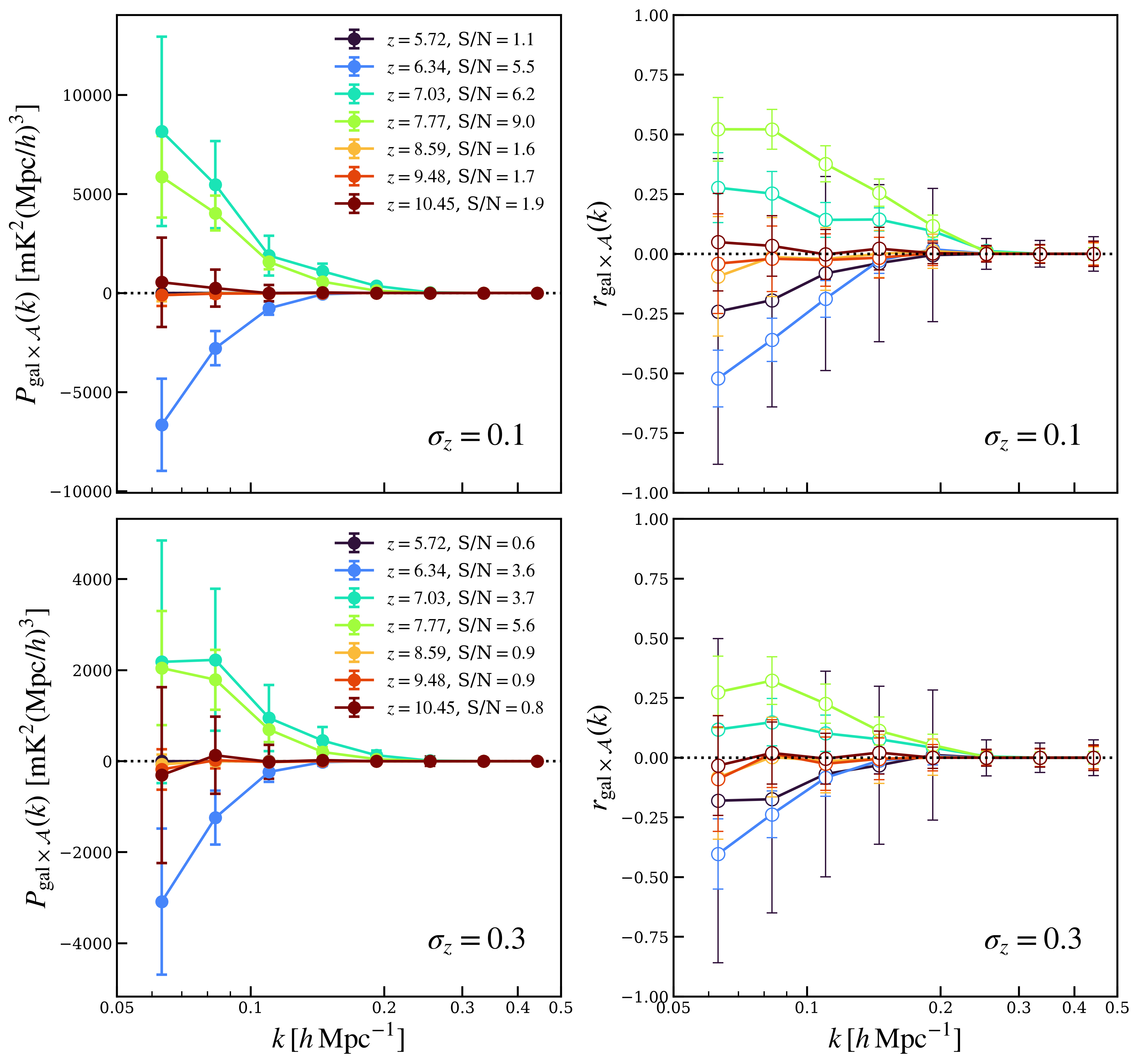}
 \caption{The same as figure~\ref{fig:constraints} but for HERA in the 331-element hexagonal configuration \citep{DillonParsons2016,DeBoer2017}.}
 \label{fig:constraints_hera}
\end{figure*}

\section{Sensitivity Forecasts for HERA} \label{sec:hera}

\begin{figure*}[!ht]
 \centering
 \includegraphics[width=\textwidth]{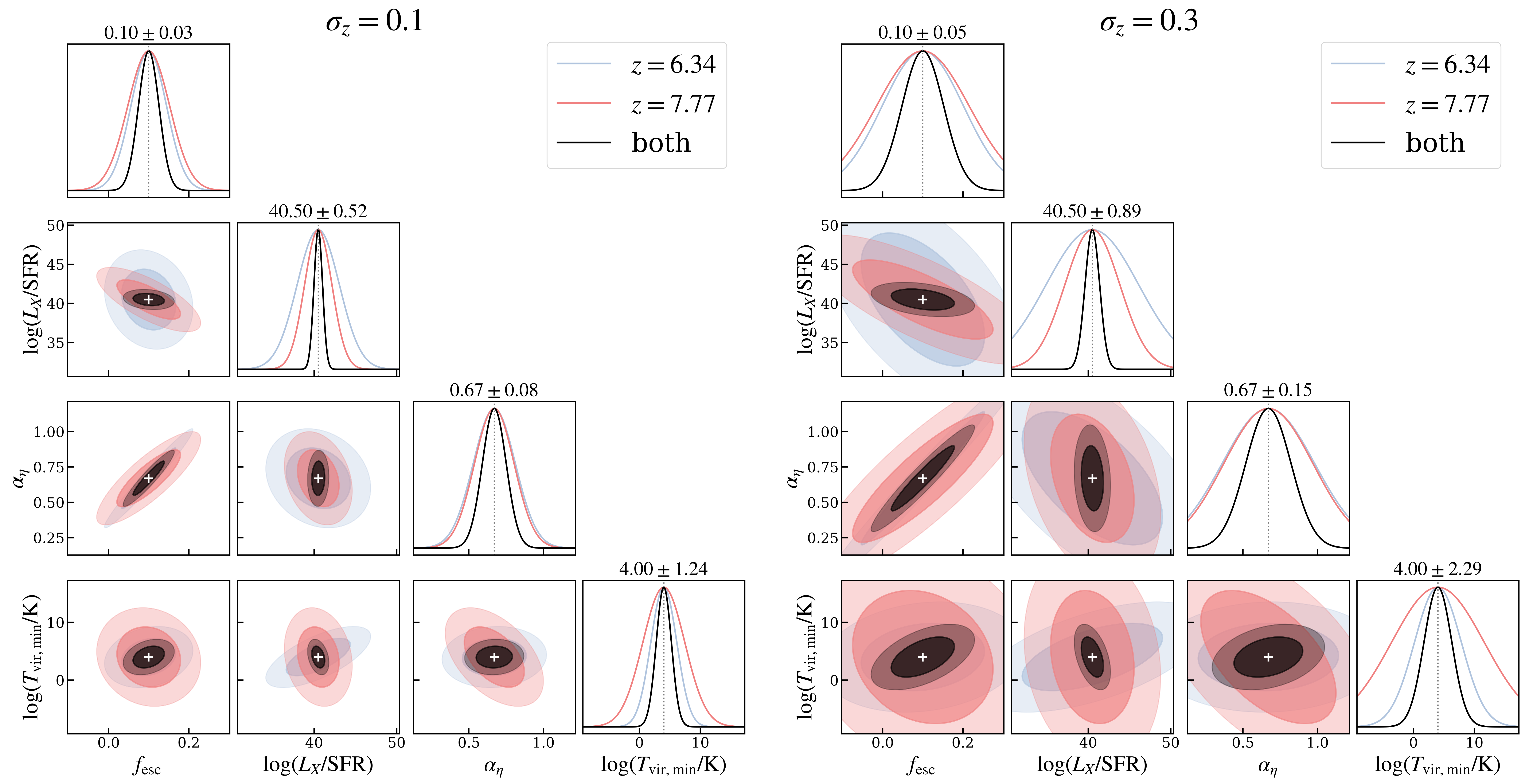}
 \caption{The same as figure~\ref{fig:fisher} but for HERA in the 331-element hexagonal configuration \citep{DillonParsons2016,DeBoer2017}.}
 \label{fig:fisher_hera}
\end{figure*}

In Section~\ref{sec:results:detect}, we evaluate the detectability of the galaxy--21\,cm$^2$ cross-power spectrum during the EoR (figure~\ref{fig:constraints}) for SKA-Low in its AA4 configuration and use Fisher matrix analysis to demonstrate its ability to constrain key physical parameters (figure~\ref{fig:fisher}). While SKA-Low represents the new generation of low-frequency interferometric 21\,cm experiments, substantial groundwork has been laid by precursor and pathfinder facilities, such as HERA \citep{DeBoer2017}, LOFAR \citep{vanHaarlem2013}, and MWA \citep{Tingay2013}, which have placed upper limits on the EoR 21\,cm power spectrum and advanced the calibration and foreground mitigation techniques \citep{Mertens2020,Trott2020,HERAI2022,Mertens2025,Nunhokee2025,HERAII2026}. In this appendix, we repeat our detectability and parameter constraint analysis using \texttt{21cmSense} for an idealized configuration of HERA, which provides a useful comparison to SKA-Low through its compact, highly redundant layout optimized for 21\,cm power spectrum measurements. 

Specifically, we consider an idealized 331-element close-packed hexagonal configuration, which should be distinguished from the currently operating HERA Phase~II array. We assume 200\,hours of HERA observations of a single field accumulated over repeated drift scans, within which an overlapping $10\,\mathrm{deg}^2$ Roman footprint is located\footnote{Note that the specified deep fields of Roman HLWAS do not overlap with the sky coverage of HERA's drift scan. The Roman--HERA synergy analyzed here is therefore hypothetical, though plausible, and is adopted to aid the comparison with SKA-Low predictions.}. Assumptions regarding the Roman galaxy fields, observing bandwidth, frequency sampling, foreground filtering, and covariance treatment are otherwise unchanged from those adopted in our SKA-Low calculations. HERA's distinct baseline sampling, thermal noise, and observing time modify the Wiener filtering, leading to different selection and weighting of 21\,cm modes when constructing $\mathcal{A}$.

Results shown by figure~\ref{fig:constraints_hera} suggest that the qualitatively positive-to-negative transition of the LBG--21\,cm$^2$ cross-correlation signal is similarly detectable by HERA but at lower statistical significance. As shown in figure~\ref{fig:fisher_hera}, combining these two epochs preserves the complementary parameter dependence found for SKA-Low, although with moderately weaker constraints. These results should be interpreted cautiously as idealized design-sensitivity forecasts. Our calculation combines idealized foreground filtering with thermal noise estimates from \texttt{21cmSense}, without realistically modeling calibration residuals, data flagging, or mutual coupling between neighboring antennas. In particular, recent HERA Phase~II analyses identify mutual coupling as the dominant residual systematic at low $k$, where it leaks foreground power into the nominal EoR window \citep{HERAII2026}. Even if this contamination is uncorrelated with the high-redshift galaxy field and therefore does not generate a mean cross-signal, it may increase the variance and thus require additional mode excision. Because $\mathcal{A}$ is constructed from pairs of surviving 21\,cm modes, such excision can also affect the mode coupling that gives rise to the cross-correlation signal. Reaching the constraining power similar to that shown in figures~\ref{fig:constraints_hera} and \ref{fig:fisher_hera} therefore requires effective control of these effects.

\section*{Acknowledgments}

GS acknowledges support from a CIERA Postdoctoral Fellowship. AL and AK were supported in part by NASA grant 80NSSC26K0183. Part of the research was carried out at the Jet Propulsion Laboratory (JPL), California Institute of Technology, under a contract with the National Aeronautics and Space Administration (80NM0018D0004). CAFG was supported by NSF through grants AST-2108230, AST-2307327 and AST-2606015; by NASA through grants 80NSSC22k0809, 80NSSC22K1124 and 80NSSC24K1224; by STScI through grant JWST-AR-03252.001-A; and by BSF through grant \#2024262. SRF was supported by NSF through award AST-2510939.  This work was performed in part at the Aspen Center for Physics, which is supported by the National Science Foundation grant PHY-2210452. ChatGPT (OpenAI; GPT-6 Astra) was used to assist with refinement of the analysis code and suggest language improvements to the manuscript. All adopted suggestions were reviewed by the authors and any resulting code changes were extensively tested. The authors take full responsibility for the research results and the final manuscript.

\bibliographystyle{JHEP}
\bibliography{paper5.bib}



\end{document}